# Enhanced Hydrogen Evolution Catalysis of Pentlandite due to the Increases in Coordination Number and Sulfur Vacancy during Cubic-Hexagonal Phase Transition

*Yuegao Liu, Chao Cai, Shengcai Zhu, Zhi Zheng, Guowu Li, Haiyan Chen, Chao Li, Haiyan Sun, I-Ming Chou, Yanan Yu, Shenghua Mei,\* and Liping Wang\**

The search for new phases is an important direction in materials science. The phase transition of sulfides results in significant changes in catalytic performance, such as $MoS_2$ and $WS_2$. Cubic pentlandite [cPn, $(Fe, Ni)_9S_8$] can be a functional material in batteries, solar cells, and catalytic fields. However, no report about the material properties of other phases of pentlandite exists. In this study, the unit-cell parameters of a new phase of pentlandite, sulfur-vacancy enriched hexagonal pentlandite (hPn), and the phase boundary between cPn and hPn are determined for the first time. Compared to cPn, the hPn shows a high coordination number, more sulfur vacancies, and high conductivity, which result in significantly higher hydrogen evolution performance of hPn than that of cPn and make the non-nano rock catalyst hPn superior to other most known nanosulfide catalysts. The increase of sulfur vacancies during phase transition provides a new approach to designing functional materials.

## 1. Introduction

The development and utilization of hydrogen energy are important ways to reduce fossil fuel dependence and achieve carbon neutrality. Platinum group metals and their derivatives play a dominant role in the $H_2$ evolution reaction (HER) and allow the fast production of $H_2$ at low overpotentials.[1–3] The low natural abundance and high price, however, impede platinum's sustainability in the hydrogen economy. Therefore, many non-noble metal HER catalysts with high HER activity have been developed to substitute Pt-based materials, such as ultrathin metallic Fe-Ni sulfide nanosheets,[4] nano Mo sulfide,[5–7] and nano $WS_2$.[8] Although these materials are very effective, the need

Y. Liu, Z. Zheng, H. Sun, I-M. Chou, S. Mei
Institute of Deep-sea Science and Engineering
Chinese Academy of Sciences
Sanya 572000, China
E-mail: mei@idsse.ac.cn

C. Cai
College of Engineering
Southern University of Science and Technology
Shenzhen 518055, China

S. Zhu
School of Materials
Sun Yat-sen University
Guangzhou 510275, China

G. Li
Crystal Structure Laboratory
Science Research Institute
China University of Geosciences (Beijing)
Beijing 100083, China

H. Chen
Mineral Physics Institute
Stony Brook University
Stony Brook, New York 11794–2100, USA

H. Chen
Argonne National Laboratory
Chicago 60439, USA

C. Li
Instrumental Analysis Center
Xi'an Jiaotong University
Xi'an 710049, China

Y. Yu
Sichuan Energy Internet Research Institute
Tsinghua University
Chengdu 610042, China

L. Wang
Academy for Advanced Interdisciplinary Studies
Southern University of Science and Technology
Shenzhen 518055, China
E-mail: wanglp3@sustech.edu.cn

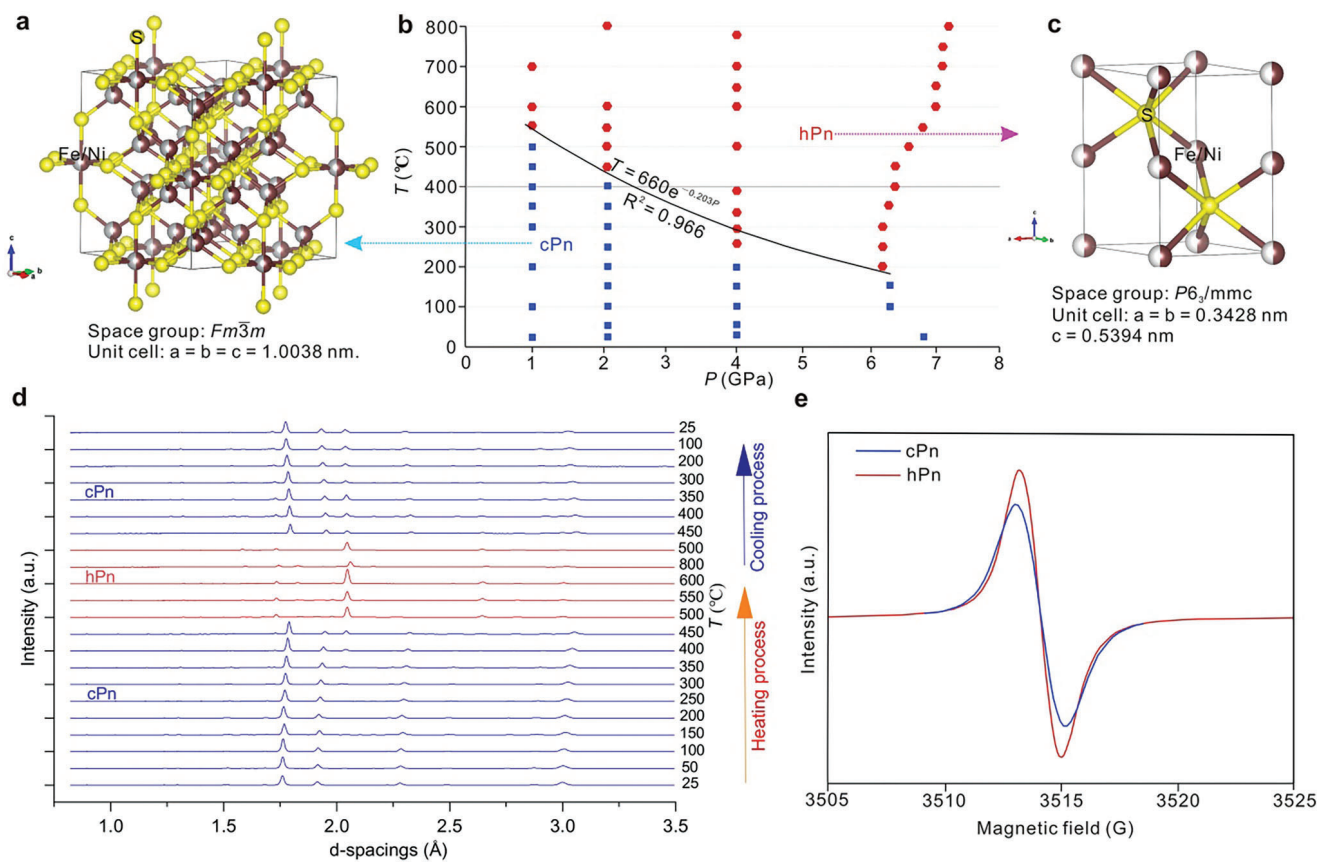

**Figure 1.** Crystal structure and phase diagram information of pentlandite. a) The crystal structure of cubic pentlandite (cPn). b) Phase diagram of pentlandite, blue squares = cPn, red hexagons = hexagonal pentlandite (hPn). c) The ideal crystal structure of hPn. d) Synchrotron X-ray characterization of the reversible phase transition of cPn to hPn at 2.1 GPa (white X-ray beam). e) Electron paramagnetic resonance (EPR) spectra of cPn and hPn.

for specific nanometer shapes required render those materials noneconomical. The natural ore pentlandite [(Fe, Ni)$_9$S$_8$], as a direct "rock" electrocatalyst without the need for further surface modifications for the HER, shows an overpotential of 280 mV at 10 mA cm$^{-2}$ under acidic conditions and high current densities (≤650 mA cm$^{-2}$) without any loss in activity for ≈170 h.[9] In addition, pentlandite is a functional material in Li-ion batteries,[10] Li-S batteries,[11] solar cells,[12] and diamond synthesis.[13] Most of the world's nickel metal comes from the natural sulfide pentlandite (Fe$_{4.5}$Ni$_{4.5}$S$_8$), which is one of the most common sulfides in magmatic nickel deposits.[14] It provides an important material foundation for the development of clean energy. The development of pentlandite's material properties can reduce environmental pollution during extracting nickel metals from sulfides. Previous studies focused on the performance of cubic pentlandite (cPn) with a symmetry group $Fm\bar{3}m$ (**Figure 1**a).[15–20] The hydrogen evolution performance of pentlandite with different compositions have been tested,[9,16,21–23] and the cubic pentlandite with a Fe:Ni ratio of 1:1 exhibits the best HER catalytic activity.[15] Former researchers have revealed the positive effects of Fe-Ni bimetallic heterostructures[24] and sulfur vacancies [19,25] on the efficient hydrogen evolution performance of pentlandite. The phase transition of some catalysts will greatly improve their catalytic performance,[26] such as the phase transitions from 2H-MoS$_2$ to 1T-MoS$_2$[6,27,28] and from 2H-WS$_2$ to 1T-WS$_2$.[8] In high-pressure experiments (1–7 GPa), we synthesized a new phase of pentlandite (Fe$_{4.5}$Ni$_{4.5}$S$_8$), sulfur-vacancy enriched hexagonal pentlandite (hPn). To date, no reports about the crystal structure and cell parameters of hexagonal pentlandite exist. In addition, the formation P–T conditions of hPn and a comparison between the HER performance of hPn and cPn were unknown.

In this study, we determined the phase boundaries between cPn and hPn based on synchrotron X-ray data. The crystal structure of hPn was determined for the first time in the world using powder and single crystal X-ray diffraction (XRD), transmission electron microscopy (TEM), and extended X-ray absorption fine structure (EXAFS). Through a comparison between the crystal structures of cPn and hPn, combined with the theoretical calculation of the density of states and the free energy of hydrogen adsorption, we found that the increase in the coordination number and sulfur vacancy during the phase transition from cubic to hexagonal pentlandite rock provided outstanding enhanced hydrogen evolution catalysis. Vacancies are fundamentally important because they are closely related to many physicochemical properties, such as electric, mechanical, thermal, optical, magnetic, and catalytic properties.[29–31] The sulfur vacancies of the catalyst rocks in this study were caused by a phase transition by changing the P–T conditions without any plasma, reduction gas, reagents, or solvent treatment. Our new hPn rock catalysts and novel method of sulfur vacancies introduction by phase transition method address some obstacles for the industrial application of hydrogen evolution catalysts and establish new avenues for designing catalysts.

## 2. Results and Discussion

### 2.1. Determination of the Phase Boundary of cPn and hPn

The molecular formula of cPn in this study is $Fe_{4.37-4.48}Ni_{4.59-4.66}Co_{0.05-0.06}S_8$ with minor Cu, Te, and Se (Table S1, Supporting Information). Here, we regarded this natural mineral as perfect pentlandite, $Fe_{4.5}Ni_{4.5}S_8$. Synchrotron radiation X-ray testing showed that the cPn-hPn phase transitions occurred at 550, 450, 258, and 200 °C and at 1.0, 2.1, 4.0, and 6.2 GPa, respectively, and the phase boundary could be represented by the following: $T$ (°C) = $660e^{-2.203P}$, where $P$ is in GPa (Figure 1b; raw data in Table S2, Supporting Information). During the step heating in the synchrotron radiation X-ray experiment at 2.1 GPa, each temperature was maintained for 5 min, and cPn was transformed to hPn at 450 °C; during the cooling process from 800 °C to room temperature with an average rate of 30 °C min$^{-1}$, hPn was reversed to cPn at the same transition temperature (450 °C) (Figure 1d); therefore, the phase transition is an enantiotropic transformation under a slow cooling process. However, if the sample was quenched from high temperature, the hPn could maintain its crystal form at room $P$–$T$.

### 2.2. Production of hPn

We applied a high $P$–$T$ environment to the cPn powder samples through a multi-anvil press (Figure S1, Supporting Information; ref. [32]); the samples were kept at 4 GPa/550 °C for 5 h, and then quenched. We obtained hPn samples with a diameter of ≈3 mm (Figure S2a, Supporting Information). The map analyses of the synthesized hexagonal pentlandite using scanning electron microscopy showed that the samples were homogeneous in Fe, Ni, and S composition and had not been separated into two or more minerals (Figure S2, Supporting Information). To prevent mineral oxidation as much as possible, both hPn and cPn were ground to 0.2–2 μm in a glove box with an agate mortar (Figures S3 and S4, Supporting Information) for XRD, TEM, EXAFS, electron paramagnetic resonance (EPR), X-ray photoelectron spectroscopy (XPS), and electrochemical measurements.

### 2.3. The Unit Cell Parameters of hPn

The accurate composition of the synthetic hPn was consistent with the X-ray photoelectron spectroscopy (XPS) analysis (Figure S5 and Table S3, Supporting Information). According to the refinement from powder XRD data (Figure S6 and Table S4, Supporting Information), the space group of the new phase pentlandite is $P6_3/mmc$ ($\alpha = \beta = 90°$, $\gamma = 120°$). The TEM images also show hexagonal features (Figure S7b,c, Supporting Information). Based on the single crystal diffraction test, the unit cell parameters of hPn are as follows: a = b = 3.428 Å and c = 5.394 Å (Figure 1c). In this crystal structure, S, Fe, and Ni are all six coordinated. Some researchers[33] believe that rhombohedral pentlandite exists with cell parameters of a = b = 0.69062 nm, c = 1.72095 nm, and V = 0.71085 nm$^3$. If the Ni/Fe-S system is a rhombohedral structure (a triangular crystal system), then Ni and Fe would have a 5-coordinate structure, such as that of Ni in NiS (CIF: 1011038).[34] However, the EXAFS data (Table S5 and Figure S8, Supporting Information) show that S, Fe, and Ni are all six coordinated, providing supporting evidence of its hexagonal crystal structure.

In the ideal crystal structure of hPn (Figure 1c), one unit cell (FeNiS$_2$) contains two S atoms and two metal atoms, and the number ratio of S to metal atoms is 1:1, which is different from that of cPn with a ratio of 8:9. Although this change in element proportion often occurs during the phase transition, we still would like to understand the imperfection of this structure. cPn is believed to have 0.275 sulfur vacancies per unit cell.[35] The S signal observed at ≈3515 G in the electron paramagnetic resonance (EPR) spectra indicates the existence of sulfur vacancies; a larger magnitude corresponds to more sulfur vacancies.[36] It is clear that hPn has more sulfur vacancies than cPn (Figure 1e). The calculation of phase transition mechanism also shows that the NiS$_6$ polyhedron is distorted and the S site in hPn is distorted or even defective (**Figure 2**). The discontinuous crystal stripes on high-resolution transmission electron microscopy (HRTEM) figure of sulfides usually indicate the presence of abundant sulfur vacancy defects.[7] The HRTEM of hPn shows clear discontinuous crystal stripes (Figure S9a,c, Supporting Information), but it is not obvious in cPn (Figure S9d, Supporting Information). Thus, some sulfur vacancies formed during the phase transition process from cPn to hPn. However, the composition of hPn barely changes (Table S1, Supporting Information); that is, the number ratio of S to metal atoms of hPn should still be 8:9. We hypothesize that four and a half unit cells (Fe$_{4.5}$Ni$_{4.5}$S$_9$) with one sulfur vacancy is the approximate structure of hPn (Fe$_{4.5}$Ni$_{4.5}$S$_8$).

Generally, the sulfur vacancies are introduced by Ar plasma,[37] hydrogen plasma,[38] nitrogen plasma,[39] Ar/H$_2$ annealing,[40] chemical reduction,[41] calcination under a nitrogen atmosphere,[42] liquid-ammonia-assisted lithiation,[36] acid treatment,[43] etching by H$_2$O$_2$,[44] and electrochemical desulfurization.[45] These methods are effective for multilayer or monolayer sulfides supported on substrates or nanosheets, but they are not needed in the rock catalyst. The sulfur vacancies of the pentlandite rocks were introduced during the phase transition by changing the $P$–$T$ conditions without any plasma, reduction gas, reagent, or solvent treatment; this provides new insights for designing electrocatalysts.

### 2.4. Potential Energy Surface (PES) of the Phase Transition

To further investigate the phase transition mechanism, we utilized the SSW pathway sampling method to explore the potential energy surface (PES) of the phase transition. In this work, the cPn and hPn have different stoichiometric ratios, namely Ni$_9$S$_8$ and Ni$_8$S$_8$ for cPn and hPn. To simplify the calculation, we delete one Ni in the 17-atoms cubic phase cell (Figure 2a) to form a defective 16-atoms cubic phase (defected cPn, Figure 2b). Then, the lowest energy barrier pathway is obtained by using the SSW pathway sampling method, which has successfully been used to resolve the phase transition mechanism of AlPO$_4$.[46] As Figure 2b,c shows, after removing one Ni atom, the defected-cPn cell is distorted while the framework is kept. By an exhaustive sampling (Figure 2c,f), we find the lowest energy barrier pathway is a two-steps path with an energy barrier 0.13 eV/f.u. (f.u. = NiS) (Figure 2g). Such a low energy barrier supports the phase transition from cPn to hPn is reasonable. The $NiS_4$ tetrahedron in the cubic phase diffuses into the adjoining octahedron interstice to form $NiS_6$, while the two $NiS_6$ polyhedrons in the cubic phase are kept. Interestingly, the S sublattice transit from fcc to hcp by shearing the subgroup lattice. As a result, the orientation relation of this path is (111)c//(001)h+[1$\bar{1}$0]c//[100]h (Note: Here c and h represent cubic and hexagonal phases, respectively). The animation of the phase transition process was listed as Movie S1 (Supporting Information). Since the Ni atom diffusion, the $NiS_6$ polyhedron would be distorted. As a result, the S site in hPn is distorted or even defective. This is consistent with the fact that the hPn shows more sulfur vacancies compared to cPn.

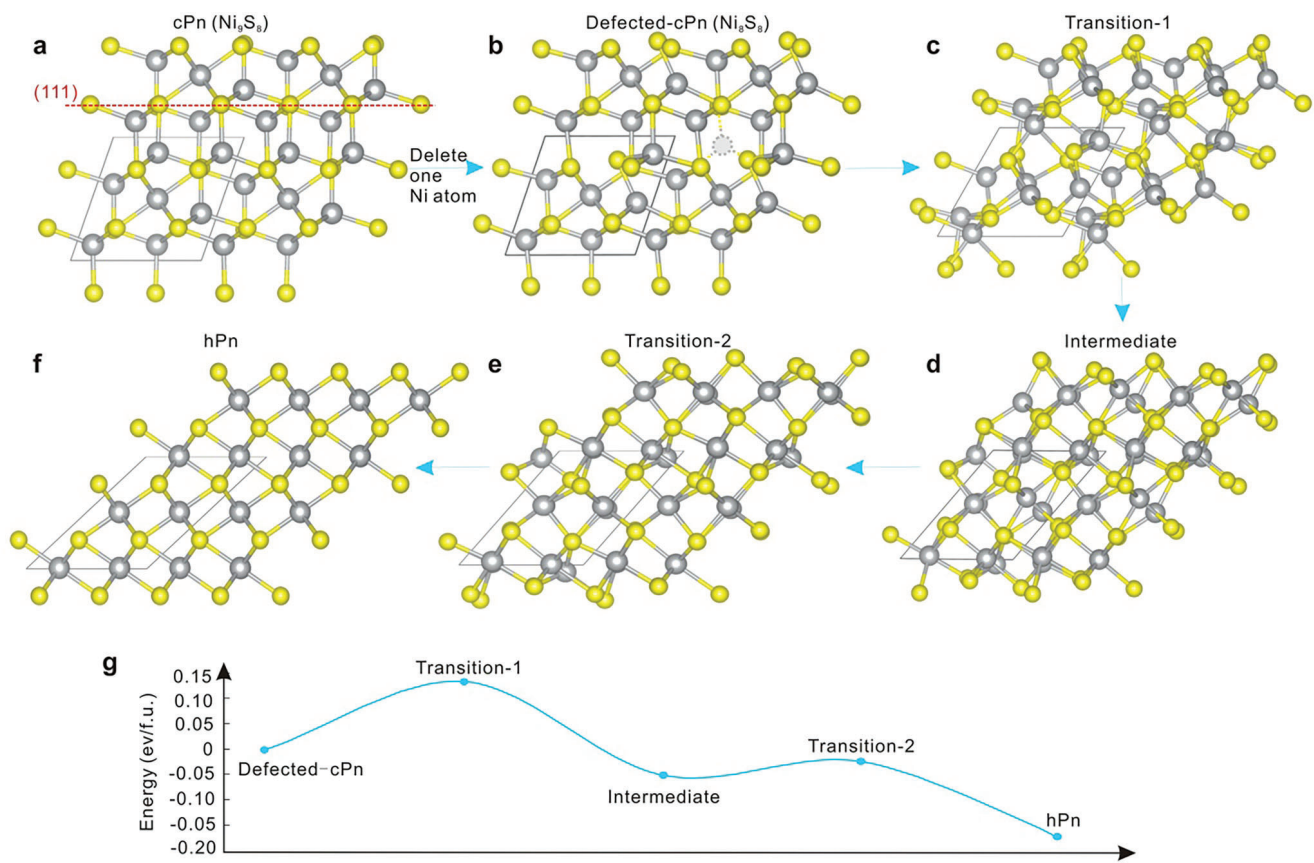

**Figure 2.** Phase transition mechanism from cPn to hPn.

## 2.5. Electrochemical Hydrogen Evolution Performance

The HER activities of cPn and hPn powders with diameters of 0.2–2 um that are grinded by hand in glove box were tested using a standard three-electrode configuration. The hPn exhibits a higher HER activity with an overpotential of −60 mV at 10 mA cm$^{−2}$, in comparison to -168 mV for cPn in 0.5 M $H_2SO_4$ solution (**Figure 3a**). The Rct of hPn in the acidic solution is 203.4 Ω, which is significantly lower than that of cPn (407.5 Ω) (Figure 3b). The Tafel slope of cPn in acidic solution is 98 mV dec$^{−1}$. This value is slightly higher than the 72 mV dec$^{−1}$ of synthetic $Ni_{4.5}Fe_{4.5}S_8$ "rocks" in ref. [9]. But the Tafel slope of hPn in the acidic solution is much lower (34 mV dec$^{−1}$) (Figure 3c), suggesting a highly promoted charge transport efficiency. In addition, hPn shows high stabilities at 300 mV and high current densities (≤205.8 mA cm$^{−2}$) without any loss in activity for ≈50 h in the acidic solution (Figure 3d), which is much higher than that of cPn (26 hrs). The electrochemical double-layer capacitance ($C_{dl}$) of the hPn (0.27 mF cm$^{−2}$) using cyclic voltammetry is much higher that of cPn (0.14 mF cm$^{−2}$) (Figure 3e). The EPR of the catalysis was invested (Figures 1e and 3f). The S signal observed at ≈3515 G indicates the existence of sulfur vacancies, and the larger the magnitude, the more sulfur vacancies are represented.[36] The hPn has more sulfur vacancies than cPn (Figure 1e). After the hydrogen evolution reaction, the sulfur vacancy of the catalysts becomes less (Figure 3f).

The overpotential of non-nano hPn rock in 0.5 M $H_2SO_4$ solution is 60 mV at 10 mA cm$^{−2}$, which is lower than that of most nano Mo sulfide (150–250),[47] $WS_2$ nanosheets (150–200 mV),[8] β-NiS nanosheets,[4] cubic pentlandite nanoparticles,[21] and so on (**Table 1**).

The Tafel slope of hPn is much lower than that of cPn, suggesting highly promoted charge transport efficiency. The Tafel slope of cPn is 98 mV dec$^{−1}$, between 38 and 116 mV dec$^{−1}$ (Figure 3c), indicating that both the Volmer step (electrochemical hydrogen adsorption, $H_3O^+ + e^- \rightarrow H_{ads}$) and Heyrovsky step (electrochemical hydrogen desorption, $H_{ads} + H_3O^+ + e^+ \rightarrow H_2 + H_2O$) are rate-determining steps. In contrast, the Tafel slope of hPn is 34 mV dec$^{-1}$, between 29 and 38 mV dec$^{-1}$ (Figure 3c), indicating that the Heyrovsky step and Tafel step (chemical desorption, $H_{ads} + H_{ads} \rightarrow H_2$) are rate-determining steps.[48,49]

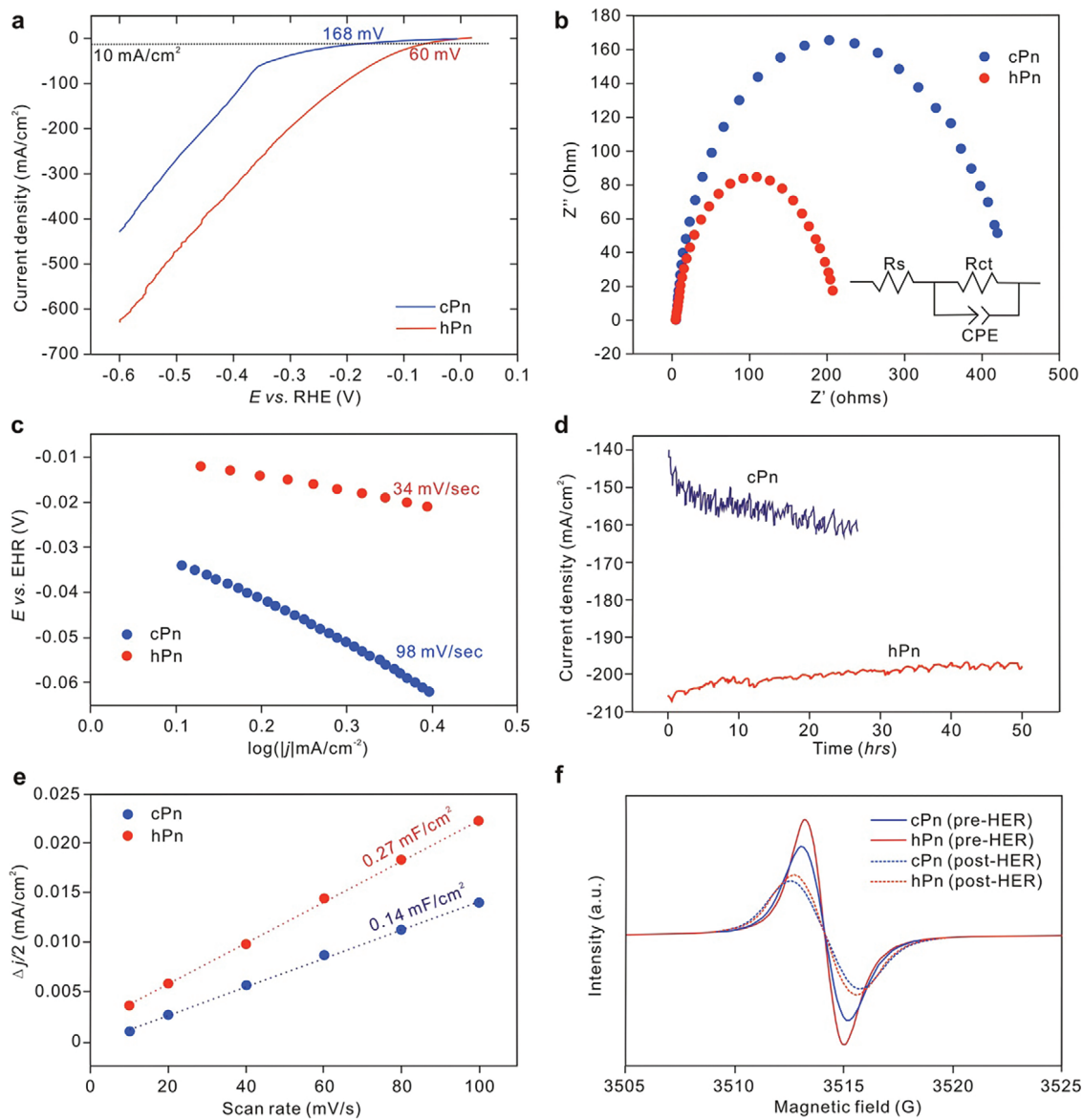

**Figure 3.** Electrochemical hydrogen evolution performance. a) The linear sweep voltammetry (LSV) curve of electrocatalysts in the acidic solution testing. b) EIS at 300 mV of cPn and hPn in the acidic solution testing. c) Tafel plot in the acidic solution. For better comparison, cPn and hPn were examined under similar conditions. d) Durable test of cPn and hPn at a constant potential of 300 mV in the acidic solution. e) The measured $C_{dl}$ values of cPn and hPn. f) EPR spectra of cPn and hPn pre-hydrogen evolution reaction (pre-HER) and post-hydrogen evolution reaction (post-HER) in the acidic (Aci) environments.

### 2.6. Calculation of the Gibbs Free Energy of Hydrogen Adsorption ($\Delta G_{H*}$)

To better understand the hydrogen evolution mechanism of cPn and hPn, we calculated the Gibbs free energy of hydrogen evolution adsorption ($\Delta G_{H*}$ value). The (111) plane is chosen to be the representation of cPn (Figure S10a, Supporting Information) since this face is considered to be one of the most efficient planes for HER.[9,24,25,51] There are two different terminals for this plane: 1) the Fe/Ni atom on the surface of this plane, called the Fe/Ni atom terminal (Figure S10b, Supporting Information), and 2) the S atom on the surface, called the S atom terminal (Figure S10d, Supporting Information). For the Fe/Ni atom terminal (slab result-1), the $\Delta G_{H*}$ values of the atoms Fe8, Fe18, and Fe1 with coordination numbers (CNs) of 3, 4, and 6, are 0.812, 0.603, and 0.316 eV, respectively; the $\Delta G_{H*}$ values of the atoms Ni4, Ni5, and Ni10 with CNs of 3, 4, and 6, are 1.337, 1.043, and 0.431 eV, respectively (**Figure 4a**; Figure S10c and Table S6, Supporting Information). Thus, the $\Delta G_{H*}$ values of Fe and Ni atoms in cPn decrease with increasing CN (Figure 4c). The $\Delta G_{H*}$ values of atoms S1 and S16 are −1.010 and 0.882 eV, respectively. The CNs of atoms S1 and S16 are 4, but the original CNs of S1 and S16 in the cPn cell are 4 and 5, respectively. Thus, for the Fe/Ni atom terminal in the (111) plane (slab result-1), a greater CN corresponds to a smaller $|\Delta G_{H*}|$ of the atoms. Compared to the Fe/Ni atom terminal (slab result-1), the CN of atoms Fe18, Fe1, Ni5, and Ni10 in the S atom terminal (slab result-2) remain the same, but the location of Fe18 and Ni5 with a CN of 4 is changed from deep layer 3 to shallow layer 2, and the location of Fe1 and Ni10 with a CN of 6 is transferred from deep layer 4 to shallow layer 3 (Figure S10c,e, Supporting Information). The $|\Delta G_{H*}|$ values of atoms Fe18, Fe1, Ni5, and Ni10 in the S atom terminal (shallower layer) are 0.720, 0.726, 1.247, and 0.737 eV, respectively (Figure 4a; Figure S10e, Supporting Information), which are much higher than those of the same atoms in the Fe/Ni atom terminal (deeper layer) (Figure 4a,d).

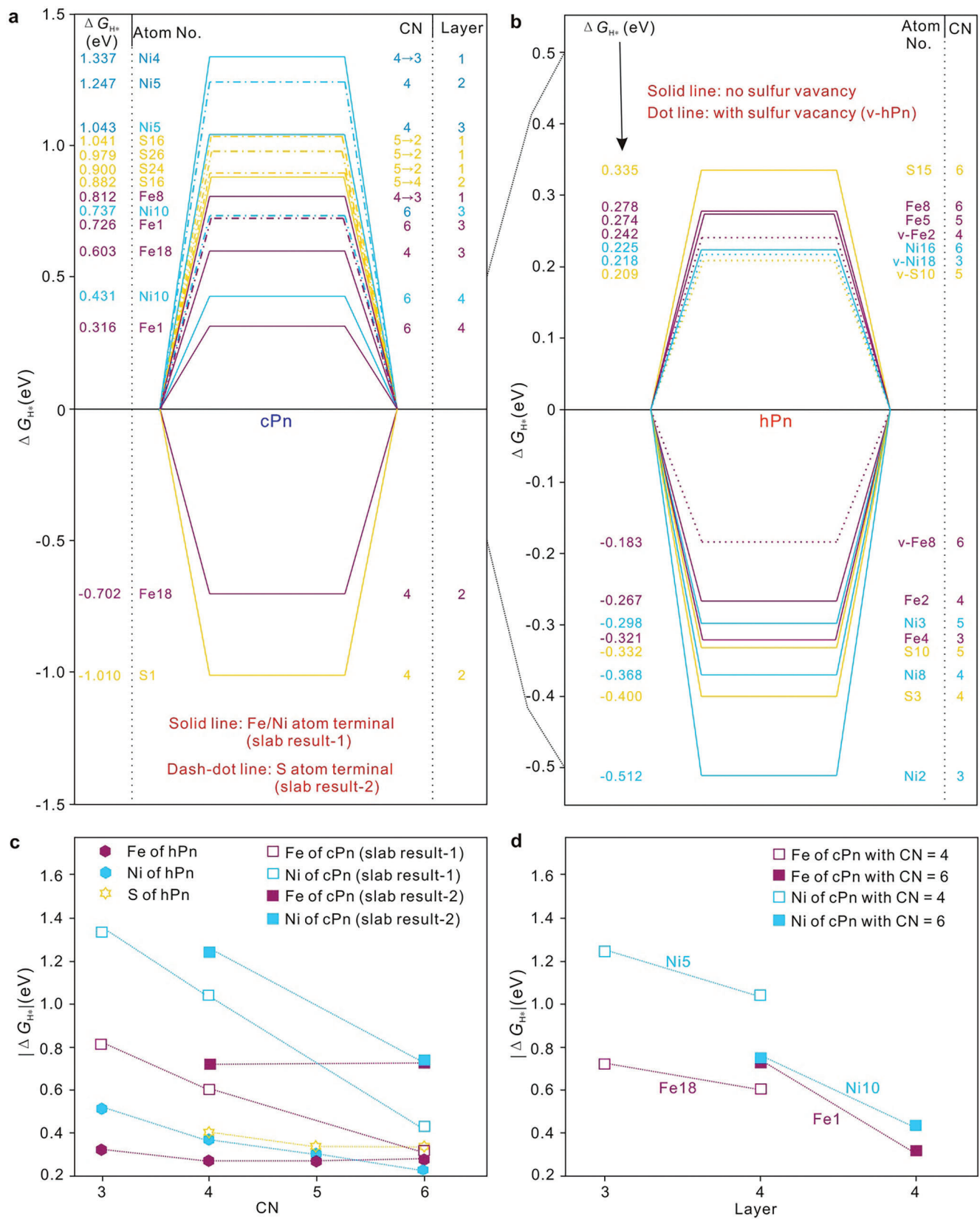

**Figure 4.** Effect of the CN, layer position, and sulfur vacancy on the $\Delta G_{H*}$. a) $\Delta G_{H*}$ on the (111) plane of cPn. b) $\Delta G_{H*}$ on the (102) plane of hPn. c) The change in $|\Delta G_{H*}|$ of different atoms in cPn with their CNs. d) The influence of the atomic layer in cPn on the $|\Delta G_{H*}|$.

**Table 1.** Comparison of hydrogen evolution performance between rock-type hPn and some outstanding nano sulfide catalysts in 0.5 M $H_2SO_4$.

| Catalyst | Overpotential at 10 mA/cm$^2$ (mV vs RHE) | Tafel slope (mV dec$^{-1}$) | Ref. |
| --- | --- | --- | --- |
| Chemical vapor deposition 2H-$MoS_2$ nanosheets | 170 | 60–98 | [6] |
| Defect-rich $MoS_2$ ultrathin nanosheets | 200 | 50 | [7] |
| Strained 1T-$WS_2$ nanosheets | 150–200 | 55 | [8] |
| β-NiS nanosheets | 202 | 48 | [4] |
| $Co_9S_8$ nanoparticles cocooned in the N-doped carbon layer | 96 | 70 | [22] |
| Silk-cocoon structured $CoS_x$ (x≈3.9, sub-10 nm) | 42 | 41 | [50] |
| Cubic $Fe_{4.5}Ni_{4.5}S_7Se$ (≈1–2 μm) | 172 | 120 | [23] |
| Cubic $Fe_{4.5}Ni_{4.5}S_6Se_2$ (≈1–2 μm) | 336 | 130 | [23] |
| $Fe_3Co_3Ni_3S_8$ nanoparticles | 517 | 139 | [21] |
| Rock-type cubic $Fe_{4.5}Ni_{4.5}S_8$ | 280 | 72 | [9] |
| Rock-type (0.2–2 μm) sulfur-vacancies enriched $P6_3/mmc$ hexagonal pentlandite ($Fe_{4.5}Ni_{4.5}S_8$) | 60 | 34 | This work |

CN = coordination number; CN = X→Y means the original coordination number of the atom in the unit cell before the slab is X, but the coordination number is changed to Y after the slab.

The (102) plane is chosen to be the representation of hPn, since this face shows the highest intensity in the XRD data (Figure S6, Supporting Information). To better understand the influence of different CNs and sulfur vacancies on the $\Delta G_{H*}$ values, four unit cells were used for the calculation (Figure S11a,b, Supporting Information). Two situations are compared: one with no sulfur vacancy (Figure S11d, Supporting Information) and the other with one sulfur vacancy (Figure S11e, Supporting Information). For the no sulfur vacancy situation, the $\Delta G_{H*}$ values of the atoms Ni2, Ni8, Ni3, and Ni6 with CNs of 3, 4, 5, and 6, are −0.512, −0.368, −0.298, and 0.225 eV, respectively (Figure 4b; Figure S11d and Table S7, Supporting Information); this shows that a greater CN corresponds to a smaller $|\Delta G_{H*}|$ of the Ni atoms (Figure 4b). The S atoms do not clearly show the above trend, but the $|\Delta G_{H*}|$ value (0.335 eV) of atom S15 (CN = 6) is much lower than that (0.400 eV) of atom S3 (CN = 4) (Figure 4b; Figure S11d and Table S7, Supporting Information). The $|\Delta G_{H*}|$ values of the Fe atom do not show much change with increasing CN. The $|\Delta G_{H*}|$ values of Fe2, Ni8, S10, and Fe8 are greatly decreased in the presence of a sulfur vacancy (Figure 4b); their values change from 0.368, 0.268, 0.332, and 0.335 eV in the no sulfur vacancy situation to 0.218, 0.242, 0.209, 0.183 eV in the presence of a sulfur vacancy, respectively (Figure 4b; Figure S11d,e and Tables S7 and S8, Supporting Information).

### 2.7. Why is the Hydrogen Evolution Performance of hPn Better Than That of cPn?

A negative $\Delta G_{H*}$ value means that $H_{ads}$ has a strong contact force with the electrode surface and is easily electrochemically adsorbed. Conversely, a positive $\Delta G_{H*}$ value indicates that the contact force between $H_{ads}$ and the electrode surface is weak, and $H_{ads}$ is not easily electrochemically adsorbed but is easily electrochemically or chemically desorbed.[52] The $\Delta G_{H*}$ values of most hPn atoms are negative, and the $\Delta G_{H*}$ values of most cPn atoms are positive (Figure 4a,b), which is consistent with the understanding of the rate-determining step.

Recent density functional theory (DFT) calculations show that unsaturated sulfur-enriched $MoS_2$ has a low hydrogen adsorption free energy and theoretically possesses the highest hydrogen generation activity.[53–55] However, this is not applicable to cPn and hPn. Overall, the peak positions and peak areas indicated by S 2p before the HER (pre-HER) of cPn and hPn are essentially the same (Figure S12a, Supporting Information). Moreover, the unsaturated sulfurs in hPn show a higher $|\Delta G_{H*}|$ than that of the saturated sulfur (Figure 4b). Thus, the unsaturated sulfur may not be the cause for the better hydrogen evolution performance of hPn than that of cPn. Here, we did not deny the importance of sulfur; our DFT calculation results show that sulfur with a CN of 2 connected with one Fe atom and one Ni atom has a lower $|\Delta G_{H*}|$ value than sulfur connected to two Fe atoms or two Ni atoms (Table S6, Supporting Information). The bridged sulfur bonds may be contributing factors in the hydrogen evolution. The zero-valent nickel can also be the active site for hydrogen evolution.[56] Both cPn and hPn show the presence existence of zero-valent nickel (Figure S13c,d, Supporting Information), which probably contributes to efficient hydrogen evolution performance of these two materials. But a more thought-provoking topic is what differences exist between these two minerals that lead to differences in hydrogen evolution performance. Hereafter, we mainly analyze the reasons from the perspective of CN and sulfur vacancy.

According to the bond-order conservation principle,[57] the lower the surface CN is, the more exotic atomic electrons can be accepted, leading to a greater adsorption energy; In contrast, a greater CN corresponds to a smaller adsorption energy. For cPn, 1/8 of the metal atoms are 6 coordinated, and the rest are 4 coordinated, while 1/4 and 3/4 sulfur atoms have CNs of 4 and 5, respectively (Figure 1a). Conversely, all atoms in hPn have a CN of 6 (Figure 1c). According to Sabatier's principle, a reactant should bind neither too strongly nor too weakly to a catalyst surface to reach optimal performance.[52] The closer the $|\Delta G_{H*}|$ value is to 0, the better the catalytic performance. Our calculation results show that a greater CN of metal atoms in cPn corresponds to a lower $|\Delta G_{H*}|$ of the atoms (Figure 4a,c). The $|\Delta G_{H*}|$ value of the Ni atoms in hPn decreases with increasing CN (Figure 4b). The S atoms do not show a clear trend in hPn, but the $|\Delta G_{H*}|$ value (0.335 eV) of atom S15 (CN = 6) is much lower than that (0.400 eV) of atom S3 (CN = 4) (Figure 4b). Overall, the increase in CN reduces the $|\Delta G_{H*}|$ value. Thus, the high coordination characteristic of the hPn atom is an important reason for its higher hydrogen evolution performance.

On the other hand, for our sample, more sulfur vacancies correspond to a better hydrogen evolution performance. Former researchers proposed one protonation mechanism in the system of pentlandite catalysts, in which sulfur vacancies are filled with hydrogen atoms preferentially.[25] This is probably one of the reasons for why the sulfur vacancy of the catalysts decreases after the hydrogen evolution reaction (Figure 3f). Many scholars have emphasized the important role of sulfur vacancies in sulfide hydrogen evolution catalysts.[25,37] Here, we analyze the influence of sulfur vacancies on hydrogen evolution mainly from the following three aspects.

I. The density of state calculation in Figure S14 (Supporting Information) shows that hPn without sulfur vacancy has higher electronic conductivity than cPn. Additionally, the sulfur vacancies increase the electronic conductivity of v-hPn (hPn with sulfur vacancies) compared to hPn without sulfur vacancies. High electronic conductivity speeds up the electron transfer rate during hydrogen evolution and improves hydrogen evolution performance. This is consistent with the fact that the Rct of hPn is significantly lower than that of cPn (Figure 3b).
II. The $|\Delta G_{H*}|$ values of atoms in hPn are greatly decreased in the presence of sulfur vacancy (Figure 4b). Thus, S vacancy lowers the H* adsorption energy of the neighboring S and metal sites, which are close to the thermos-neutral states, leading to enhanced HER activity.
III. Both cPn and hPn show that the $|\Delta G_{H*}|$ values of metal atoms in deeper layers are generally lower than those in shallower layers (Figure 4b,d), which possibly causes a strong hydrogen evolution performance in pentlandite, as a non-nano rock catalyst. In the absence of sulfur vacancies, if the metal atoms are too far from the reaction interface, hydrogen can hardly be adsorbed onto the metal surface across many atoms. Sulfur vacancies possibly alleviate this problem and expose more metal-active sites in the deeper layers. Thus, the above three aspects are potential reasons for the better catalytic performance of hPn than that of cPn.

## 3. Conclusion

The crystal structure and cell parameters of a new phase of pentlandite [(Fe,Ni)$_9$S$_8$], sulfur-vacancies enriched $P6_3$/mmc hexagonal pentlandite, which can be used in supercapacitors, battery electrodes, solar cell, and catalysis (electrocatalytic hydrogen evolution and diamond synthesis), are determined for the first time. The space group of the hexagonal pentlandite is $P6_3$/mmc ($\alpha = \beta = 90°$, $\gamma = 120°$); and the unit cell parameters are as follows: a = b = 3.428 Å and c = 5.394 Å. The $P$-$T$ environment for the stable existence of hexagonal pentlandite is also determined for the first time, and its phase boundary with cubic pentlandite conforms to the following formula: $T$ (°C) = 660e$^{-2.203P}$, where $P$ is in GPa. The high coordination number, more sulfur vacancies, and high conductivity cause the non-nano rock catalyst hexagonal pentlandite superior to other known nanosulfide catalysts. The sulfur vacancies of the pentlandite were introduced during the phase transition process just by changing $P$–$T$ conditions without any plasma, reduction gas, reagents, or solvent treatment. This method of sulfur vacancies introduction address some obstacles for the industrial application of hydrogen evolution catalysts and establish new avenues for designing electrocatalysts. In summary, this researcher provides sulfur-vacancies enriched $P6_3$/mmc hexagonal pentlandite as a new mineral material for the energy industry and catalytic field.

## 4. Experimental Section

*Synchrotron X-ray Testing at High Temperature and High Pressure at the 6BM-B hutch at Argonne National Laboratory*: The sample material was prepared by grinding pure natural cubic pentlandite to a mean grain size of ≈2 μm, which was optimal for the X-ray diffraction measurements. X-ray diffraction patterns of this material confirmed that it was a pure cubic pentlandite phase (by fitting the peaks with the "Match" program from Crystal Impact). The cubic pentlandite powder was hand-packed into the cylindrical space (2.06 mm diameter × 2.06 mm long) within a multi-anvil cell assembly (6.18 × 6.18 × 6.18 mm$^3$) to produce the lower halves of the compound test samples. NaCl plus boron nitride (BN) (NaCl: BN = 10:1 by weight), with the same sample dimensions, was also prepared and inserted into the upper half of the cylindrical space within the same sample assembly as a pressure calibration material for all four experiments. A horizontal white X-ray beam was projected through the sample assembly, perpendicular to the cylindrical axis of the cell. During the X-ray test, the test times of the pressure calibration material and the target sample were 60 s and 300 s, respectively.

Four compression experiments were performed using the compound sample assembly described above at 1, 2.1, 4, and 7 GPa, respectively, and temperature ranges from room temperature to 800 °C. The experiments' ID included Run 2584 on June 13th, 2018, and 252 453 on March 16th, 2022 (Proposal ID: 75643). To investigate whether the phase transition was reversible, the sample was gradually heated from room temperature to 800 °C at 2.1 GPa, then gradually cooled from 800 °C to room temperature at an average rate of 30°C min$^{-1}$, and measured using XRD at approximately 50 °C intervals near the phase transition point or at ≈100 °C intervals far away from the phase transition point. The raw XRD data were calibrated by Plot85 software (www.mpi.stonybrook.edu/NSLS/X17B2/Software/software.htm). Only the DvsI9 data were used to determine the phase boundary.

*Sample Preparation for Electrochemical Measurements*: Cubic pentlandite with a particle size of 0.5–1 mm and hexagonal pentlandite samples in the form of 2–3 mm sulfide blocks were ground in a glove box with an agate mortar (Figure S3, Supporting Information). The oxygen content in the glove box was not higher than 0.6%, and the grinding time was 2.5 h. The ground samples were placed in a plastic centrifuge tube in the glove box, and then the plastic centrifuge tube containing the sample and the Mitsubishi anaerobic gas production bag was put together in a sealed 15 × 30 cm plastic bag (Figure S3, Supporting Information) produced by Mitsubishi Gas Chemical Company, Inc. To avoid the oxidation of the natural samples as much as possible, the plastic bag was kept sealed until the experiments were performed; SEM, XRD, EXAFS, XPS, EPR, and electrochemical measurements were carried out immediately upon unsealing the samples.

*SEM Test*: The SEM test was performed on the Apreo C LoVac scan electron microscope at the Institute of Deep-sea Science and Engineering, Chinese Academy of Sciences (IDSSE). The energy spectrum model was AZtec X Max 50, and the testing conditions were 15 kV and 0.4 nA.

*Powder XRD Test*: The XRD test was performed using a PANalytical Empyrean powder diffractometer (model: D8 VENTURE) at IDSSE. The anode material was Cu, with K-Alpha1, K-Alpha1, and K-Beta values of 1.54060, 1.54443, and 1.39225 Å, respectively. The current and voltage during the test were 40 mA and 45 kV, respectively. The range of $2\theta$ were 20–80°. The step size was 0.013°, and the test times for cPn and hPn were 165 min and 500 min, respectively. The powder XRD data are shown in Table S4 (Supporting Information).

*Lorentz Transmission Electron Microscopy (TEM) Testing*: The TEM specimens were prepared by a focused ion beam (FIB, JEOL JIB-4000PLUS). Because the samples have weak magnetism, a Lorentz TEM ThermoFisher Talos-F200X at Xi'an Jiaotong University was used to test the structure. Figure S7 (Supporting Information) shows the selected area electron diffraction patterns of cPn (Figure S7a, Supporting Information) and hPn (Figure S7b,c, Supporting Information).

*Single Crystal Diffraction*: The single crystal X-ray diffraction data for the hexagonal pentlandite were collected by a Rigaku-Oxford diffraction XtaLAB PRO-007HF single crystal diffractometer equipped with a rotating anode micro-focus X-ray source (1.2 kW Mo$_{K\alpha}$ $\lambda = 0.71073$ Å) and a hybrid pixel array detector at China University of Geosciences (Beijing). The reflections obtained were from one single crystal fragment of excellent quality 0.15× 0.01 × 0.01 mm in size, with –4 <$h$< 4, –4 <$k$< 4, –6 <$l$< 7 with fair $R_{int}$ = 0.085. The intensity data were subjected to PL and multiscan absorption corrections. The space group was obtained by conducting the system extinction statistics and was determined to be $P6_3/mmc$. The crystal structure determination and refinement were performed using OLEX2-1.3,[58] with the SHELXT.[59] The crystal structure was determined using direct methods, which found the Fe and S atoms.

*Extended X-ray Absorption Fine Structure (EXAFS) Measurements*: The extended X-ray absorption fine structures (EXAFS) of the sample at the Ni K-edge and Fe K-edge were collected using the beamline of TPS44A1 at the National Synchrotron Radiation Research Center (NSRRC) in Taiwan. The electron energy was 3.0 GeV, and a pair of channel-cut Si (111) crystals was used in the monochromator. The acquired EXAFS data were processed according to standard procedures using the ATHENA module implemented in the IFEFFIT software packages.[60] The k3-weighted EXAFS spectra were obtained by subtracting the post-edge background from the overall absorption and then normalizing with respect to the edge-jump step. Subsequently, k2-weighted $\chi$(k) data of the Ni K-edge and Fe K-edge were Fourier transformed to real (R) space (dk = 1.0 Å$^{-1}$) to separate the EXAFS contributions from different coordination shells.[61] To obtain the quantitative structural parameters around central atoms, a least-squares curve parameter fitting was performed using the ARTEMIS module of IFEFFIT software packages.[62]

*The Calculation of the Potential Energy Surface (PES) of Phase Transition*: In this work, the Stochastic Surface Walking (SSW) pathway sampling method[63] to explore the potential energy surface (PES) of the NiS (the defect Ni$_9$S$_8$) phase transition was utilized. In the SSW method, the movement on the potential surface was guided by the random soft mode (second derivative) direction, which was capable of exploring potential energy surface exhaustively and unbiasedly. Now, the SSW was a module implemented in the package of Large-scale Atomic Simulation with Neural Network Potential (LASP, webpage www.lasphub.com).[64] In this work, the pathway sampling was carried out in a 16-atom cell, and more than 100 pairs of initial state/final state (IS/FS) were collected at the density functional theory (DFT) level. Based on the large IS/FS pairs datasets, the pathways connecting the initial structure and the final structure could be determined in an atom-to-atom correspondence, then the transition state could be seamlessly located by using the variable-cell double-ended surface walking approach (VC-DESW).[63] After transition state searching, the lowest energy barrier pathways could be determined by sorting the energy barriers. This method has successfully been used to predict the low-energy pathways of AlPO$_4$.[46] All calculations were performed using the plane wave DFT program, Vienna ab initio simulation package (VASP)[65] where the electron-ion interaction of Ni and S atoms were represented by the projector augmented wave (PAW) scheme,[66] and the exchange-correlation functional utilized was GGA-PBE.[67] In the SSW pathway sampling, the following setups was adopted to speed up the PES exploration: the kinetic energy plane wave cutoff 400 eV; the Monkhorst-Pack k-point mesh of (3×3×3) set for 16 and 17-atom supercell.[68] In the transition state search: the plane-wave cutoff 500 eV, Monkhorst-Pack k-point mesh (5×5×5). For all the structures, both lattice and atomic positions were fully optimized until the maximal stress component was below 0.1 GPa and the maximal force component below 0.01 eV Å$^{-1}$.

*X-Ray Photoelectron Spectroscopy Measurements*: X-ray photoelectron spectroscopy measurements were conducted using ESCALAB 220i-XL XPS (Thermo Scientific, UK) at Shenzhen University. Monochromated Al K$\alpha$ X-rays at 1486.6 eV were used. The target voltage and current were 15 kV and 10 mA, respectively. The vacuum chamber pressure was less than 2 × 10$^{-6}$ Pa, and the X-ray spot was 650 µm in diameter. The pass energy was 30 eV, and the measurement step was 0.1 eV. The C peak standard was 284.8 eV. Data analysis was carried out using XPSPEAK4.1 software and Origin software Version 2018. The Shirley model was used for background subtraction when fitting photoelectron spectra. The XPS results are presented in Table S9 (Supporting Information).

*Electron Paramagnetic Resonance Testing*: Electron paramagnetic resonance (EPR) was performed at room temperature using a Bruker ER 200D spectrometer instrument at Tsinghua University. Twenty-eight mg powder samples were placed in a quartz tube, and then the quartz tube was placed into an electron spin resonance tester. The microwave frequency was used to maintain the X-band range from 9.8591 to 9.8599 G Hz, while the microwave energy was kept at ≈20 mW to avoid saturation. The raw EPR data are shown in Table S10 (Supporting Information).

*Electrochemical Measurements*: The electrochemical performance of the catalyst was measured via a standard three-electrode configuration using a CHI760e potentiostat at the School of Physics and Electronics, Central South University. The catalyst materials supported on carbon cloth (ink of 2 mg mL$^{-1}$ in ethanol/isopropanol (volume ratio of 3:1), 4 µL ink was added dropwise onto the carbon cloth), Ag/AgCl, and a graphite rod were used as the working electrode, reference electrode, and counter electrode, respectively. The geometric area of the working electrode was 0.0706 cm$^2$. 0.5 MH$_2$SO$_4$ solution was used as the electrolyte for alkaline and acidic measurements, respectively. Linear sweep voltammetry measurements were conducted at a 10 mV s$^{-1}$ scan rate. The EIS was recorded from 100 000 to 0.01 Hz. All measurements were conducted at room temperature. The raw electrochemical testing data are listed in Table S11 (Supporting Information).

*Calculation of Density of State and the Gibbs Free Energy of Hydrogen Adsorption*: To better assess the findings on the activity of cPn and hPn and to understand their origins, the hydrogen adsorption energy $\Delta E_H$ was calculated. The unit cell of cubic pentlandite was a = b = c = 1.0038 nm (CIF: 9 004 076). Theoretical calculations were implemented in the Vienna ab initio Simulation Package (VASP)[65,69] using the plane-wave based density functional theory (DFT) method. The generalized gradient approximation (GGA) of Perdew-Burke-Ernzerhof (PBE)[67] was applied for the exchange-correlation functional. A vacuum space of 15 Å in the z-axis direction was used in these calculations. The k-points of 2 × 2 × 1 were automatically generated with gamma symmetry. Spin polarization was considered in all the calculations. The cutoff energy was set as 500 eV and the electronic energy convergence criterion was set at $10^{-5}$ eV. The Gibbs free energy difference ($\Delta G_{H^*}$) in the water redox reactions was defined as follows: $\Delta G_{H^*} = \Delta E + \Delta E_{ZPE} - T\Delta S$ where $\Delta E$, $\Delta E_{ZPE}$, and $\Delta S$ represent the energy difference of adsorption, zero-point energy, and entropy between the adsorbed state and the corresponding freestanding state, respectively. Herein, the values for $\Delta E_{ZPE}$ and $\Delta S$ were obtained via DFT calculations and standard thermodynamic data.

## Supporting Information

Supporting Information is available from the Wiley Online Library or from the author.

## Acknowledgements

Y.L., C.C., S.Z., and Z.Z. contributed equally to this work. This work was supported by the Hainan Provincial Joint Project of Sanya Yazhou Bay Science and Technology City (2021CXLH0027), the New Star of South China Sea Talent Project (NHXXRCXM202339), the funds from Chinese Academy of Sciences(Grants No. QYZDY-SSW-DQC029 and XDA22040501), the Central Guidance on Local Science and Technology Development Fund (ZY2021HN15), the Major Science and Technology Infrastructure Project of Material Genome Big-science Facilities Platform from the Municipal Development and Reform Commission of Shenzhen, and the National Natural Science Foundation of China(No. 41973055 and No. 42130109). Portions of this work were performed at 6BM-B, APS, ANL. The APS is supported by the DOE-BES, under Contract No. DE-AC02-06CH11357. The authors thank the Taiwan Light Source for the help in the EXAFS test. Dr. Xuegong Li, Jie Dai, Rui Zhao, and Jin Lin helped a lot during sample preparation.

## Conflict of Interest

The authors declare no conflict of interest.

## Data Availability Statement

The data that support the findings of this study are available in the supplementary material of this article.

[1] L. Huang, P. C. Chen, M. Liu, X. Fu, P. Gordiichuk, Y. Yu, C. Wolverton, Y. Kang, C. A. Mirkin, *Proc. Natl. Acad. Sci. USA* **2018**, *115*, 3764.
[2] F. Lv, B. Huang, J. Feng, W. Zhang, K. Wang, N. Li, J. Zhou, P. Zhou, W. Yang, Y. Du, *Natl. Sci. Rev.* **2021**, *8*, nwab019.
[3] J. Wang, J. Yu, J. Wang, K. Wang, L. Yu, C. Zhu, K. Gao, Z. Gong, Z. Li, R. Devasenathipathy, D. Cai, H. Xie, G. Lu, *Small* **2023**, *19*, 2207135.
[4] X. Long, G. Li, Z. Wang, H. Zhu, T. Zhang, S. Xiao, W. Guo, S. Yang, *J. Am. Chem. Soc.* **2015**, *137*, 11900.
[5] H. I. Karunadasa, E. Montalvo, Y. Sun, M. Majda, J. R. Long, C. J. Chang, *Science* **2012**, *335*, 698.
[6] D. Voiry, R. Fullon, J. Yang, C. de Carvalho Castro e Silva, R. Kappera, I. Bozkurt, D. Kaplan, M. J. Lagos, P. E. Batson, G. Gupta, *Nat. Mater.* **2016**, *15*, 1003.
[7] J. Xie, H. Zhang, S. Li, R. Wang, X. Sun, M. Zhou, J. Zhou, X. W. Lou, Y. Xie, *Adv. Mater.* **2013**, *25*, 5807.
[8] D. Voiry, H. Yamaguchi, J. Li, R. Silva, D. C. Alves, T. Fujita, M. Chen, T. Asefa, V. B. Shenoy, G. Eda, *Nat. Mater.* **2013**, *12*, 850.
[9] B. Konkena, K. J. Puring, I. Sinev, S. Piontek, O. Khavryuchenko, J. P. Dürholt, R. Schmid, H. Tüysüz, M. Muhler, W. Schuhmann, U.-P. Apfel, *Nat. Commun.* **2016**, *7*, 12269.
[10] Y. Zhou, D. Yan, H. Xu, J. Feng, X. Jiang, J. Yue, J. Yang, Y. Qian, *Nano Energy* **2015**, *12*, 528.
[11] C. Dai, J. M. Lim, M. Wang, L. Hu, Y. Chen, Z. Chen, H. Chen, S. J. Bao, B. Shen, Y. Li, *Adv. Funct. Mater.* **2018**, *28*, 1704443.
[12] S. H. Chang, M. D. Lu, Y. L. Tung, H. Y. Tuan, *ACS Nano* **2013**, *7*, 9443.
[13] Y. N. Palyanov, Y. M. Borzdov, Y. V. Bataleva, I. N. Kupriyanov, *Diamond Relat. Mater.* **2021**, *120*, 108660.
[14] Y. Liu, W. Y. Li, X. B. Lü, Y. R. Liu, B. X. Ruan, X. Liu, *Ore. Geol. Rev.* **2017**, *91*, 419.
[15] S. Piontek, C. Andronescu, A. Zaichenko, B. Konkena, J. P. Kai, B. Marler, H. Antoni, I. Sinev, M. Muhler, D. Mollenhauer, B. R. Cuenya, W. Schuhmann, U. P. Apfel, *ACS Catal.* **2018**, *8*, 987.
[16] M. Smialkowski, D. Siegmund, K. Stier, L. Hensgen, M. P. Checinski, U.-P. Apfel, *ACS Mater. Au* **2022**, *2*, 474.
[17] C. Zhang, Y. Cui, Y. Yang, L. Lu, S. Yu, Z. Meng, Y. Wu, Y. Li, Y. Wang, H. Tian, *Adv. Funct. Mater.* **2021**, *31*, 2105372.
[18] C. Zhang, H. Nan, H. Tian, W. Zheng, *J. Alloys Compd.* **2020**, *838*, 155685.
[19] Y. Wu, Y. Yang, Y. Li, C. Zhang, Y. Wang, H. Tian, *Ceram. Int.* **2021**, *47*, 12002.
[20] C. Zhang, C. Jiang, Q. Tang, Z. Meng, Y. Li, Y. Wang, Y. Cui, W. Shi, S. Yu, H. Tian, *J. Energy Chem.* **2023**, *78*, 438.
[21] M. B. Z. Hegazy, K. Harrath, D. Tetzlaff, M. Smialkowski, D. Siegmund, J. Li, R. Cao, U.-P. Apfel, *iScience* **2022**, *25*, 105148.
[22] J. Mujtaba, L. He, H. Zhu, Z. Xiao, G. Huang, A. A. Solovev, Y. Mei, *ACS Appl. Nano Mater.* **2021**, *4*, 1776.
[23] M. Smialkowski, D. Siegmund, K. Pellumbi, L. Hensgen, H. Antoni, M. Muhler, U.-P. Apfel, *Chem. Commun.* **2019**, *55*, 8792.
[24] L. Lu, S. Yu, H. Tian, *J. Colloid Interface Sci.* **2022**, *607*, 645.
[25] I. Zegkinoglou, A. Zendegani, I. Sinev, S. Kunze, H. Mistry, H. S. Jeon, J. Zhao, M. Y. Hu, E. E. Alp, S. Piontek, *J. Am. Chem. Soc.* **2017**, *139*, 14360.
[26] J. Kim, H. Kim, B. Ruqia, M. J. Kim, Y. J. Jang, T. H. Jo, H. Baik, H. S. Oh, H. S. Chung, K. Baek, *Adv. Mater.* **2021**, *33*, 2105248.
[27] M. A. Lukowski, A. S. Daniel, F. Meng, A. Forticaux, L. Li, S. Jin, *J. Am. Chem. Soc.* **2013**, *135*, 10274.
[28] H. Wang, Z. Lu, S. Xu, D. Kong, J. J. Cha, G. Zheng, P. C. Hsu, K. Yan, D. Bradshaw, F. B. Prinz, *Proc. Natl. Acad. Sci. USA* **2013**, *110*, 19701.
[29] Z. Sun, K. Ikemoto, T. M. Fukunaga, T. Koretsune, R. Arita, S. Sato, H. Isobe, *Science* **2019**, *363*, 151.
[30] Z. Zhang, G. Liu, X. Cui, Y. Gong, Y. Yi, Q. Zhang, C. Zhu, F. Saleem, B. Chen, Z. Lai, *Sci. Adv.* **2021**, *7*, eabd6647.
[31] H. Liang, M. Jiao, Y. Huang, P. Yu, X. Ye, Y. Wang, Y. Xie, Y.-F. Cai, X. Rong, J. Du, *Natl. Sci. Rev.* **2023**, *10*, nwac262.
[32] Y. Liu, I. M. Chou, J. Chen, N. Wu, W. Li, L. Bagas, M. Ren, Z. Liu, S. Mei, L. Wang, *Natl. Sci. Rev.* **2023**, *10*, nwad159.
[33] H. F. Liu, G. W. Li, Q. Lu, N. C. Shi, Z. D. Tang, P. Xiao, *Earth Sci.* **2012**, *37*, 501.
[34] N. Alsén, *Geologiska Föreningen i Stockholm Förhandlingar* **1925**, *47*, 19.
[35] O. Knop, M. A. Ibrahim, *Can. J. Chem.* **1961**, *39*, 297.
[36] Y. Yin, J. Han, Y. Zhang, X. Zhang, P. Xu, Q. Yuan, L. Samad, X. Wang, Y. Wang, Z. Zhang, *J. Am. Chem. Soc.* **2016**, *138*, 7965.

[37] H. Li, C. Tsai, A. L. Koh, L. Cai, A. W. Contryman, A. H. Fragapane, J. Zhao, H. S. Han, H. C. Manoharan, F. Abild-Pedersen, J. K. Nørskov, X. Zheng, *Nat. Mater.* **2016**, *15*, 48.
[38] A. Y. Lu, X. Yang, C. C. Tseng, S. Min, S. H. Lin, C. L. Hsu, H. Li, H. Idriss, J. L. Kuo, K. W. Huang, *Small* **2016**, *12*, 5530.
[39] J. Jiang, Q. Zhang, A. Wang, Y. Zhang, F. Meng, C. Zhang, X. Feng, Y. Feng, L. Gu, H. Liu, L. Han, *Small* **2019**, *15*, 1901791.
[40] Y. Li, C. Jiang, Y. Yang, C. Zhang, J. Xu, Y. Zeng, S. Yu, H. Tian, W. Zheng, *Appl. Surf. Sci.* **2022**, *604*, 154470.
[41] X. Dai, X. Wang, G. Lv, Z. Wu, Y. Liu, J. Sun, Y. Liu, Y. Chen, *Small* **2023**, *19*, 2302267.
[42] H. Jing, G. Xu, B. Yao, J. Ren, Y. Wang, Z. Fang, Q. Liang, R. Wu, S. Wei, *ACS Appl. Energy Mater.* **2022**, *5*, 10187.
[43] A. Li, H. Pang, P. Li, N. Zhang, G. Chen, X. Meng, M. Liu, X. Liu, R. Ma, J. Ye, *Appl. Catal., B* **2021**, *288*, 119976.
[44] X. Wang, Y. Zhang, H. Si, Q. Zhang, J. Wu, L. Gao, X. Wei, Y. Sun, Q. Liao, Z. Zhang, *J. Am. Chem. Soc.* **2020**, *142*, 4298.
[45] C. Tsai, H. Li, S. Park, J. Park, H. S. Han, J. K. Nørskov, X. Zheng, F. Abild-Pedersen, *Nat. Commun.* **2017**, *8*, 15113.
[46] S. C. Zhu, G. W. Chen, D. Zhang, L. Xu, Z. P. Liu, H. k. Mao, Q. Hu, *J. Am. Chem. Soc.* **2022**, *144*, 7414.
[47] J. Kibsgaard, Z. Chen, B. N. Reinecke, T. F. Jaramillo, *Nat. Mater.* **2012**, *11*, 963.
[48] J. M. Bockris, E. Potter, *J. Electrochem. Soc.* **1952**, *99*, 169.
[49] T. Shinagawa, A. T. Garcia-Esparza, K. Takanabe, *Sci. Rep.* **2015**, *5*, 13801.
[50] C. Wang, T. Wang, J. Liu, Y. Zhou, D. Yu, F. Han, Q. Li, J. Chen, Y. Huang, *Energy Environ. Sci.* **2018**, *11*, 2467.
[51] C. L. Bentley, C. Andronescu, M. Smialkowski, M. Kang, T. Tarnev, B. Marler, P. R. Unwin, U. P. Apfel, W. Schuhmann, *Angew. Chem., Int. Ed.* **2018**, *57*, 4093.
[52] P. Sabatier, *Ber. Dtsch. Chem. Ges.* **1911**, *44*, 1984.
[53] B. Hinnemann, P. G. Moses, J. Bonde, K. P. Jørgensen, J. H. Nielsen, S. Horch, I. Chorkendorff, J. K. Nørskov, *J. Am. Chem. Soc.* **2005**, *127*, 5308.
[54] Q. Xu, Y. Liu, H. Jiang, Y. Hu, H. Liu, C. Li, *Adv. Energy Mater.* **2019**, *9*, 1802553.
[55] T. F. Jaramillo, K. P. Jørgensen, J. Bonde, J. H. Nielsen, S. Horch, I. Chorkendorff, *Science* **2007**, *317*, 100.
[56] C. Hu, Q. Ma, S. Hung, Z. Chen, D. Ou, B. Ren, H. M. Chen, G. Fu, N. Zheng, *Chem* **2017**, *3*, 122.
[57] E. Shustorovich, *Surf. Sci. Rep.* **1986**, *6*, 1.
[58] O. V. Dolomanov, L. J. Bourhis, R. J. Gildea, J. A. Howard, H. Puschmann, *J. Appl. Crystallogr.* **2009**, *42*, 339.
[59] G. M. Sheldrick, *Acta Crystallogr., Sect. C: Struct. Chem.* **2015**, *71*, 3.
[60] B. Ravel, M. Newville, *J. Synchrotron Radiat.* **2005**, *12*, 537.
[61] R. Prins, D. Koningsberger, *X-Ray Absorption: Principles, Applications, Techniques of EXAFS, SEXAFS, and XANES*, (eds D. C. Koningsberger, R. Prins), Wiley, Hoboken, NJ **1988**, 92.
[62] J. J. Rehr, R. C. Albers, *Rev. Mod. Phys.* **2000**, *72*, 621.
[63] X. J. Zhang, Z. P. Liu, *J. Chem. Theory Comput.* **2015**, *11*, 4885.
[64] S. D. Huang, C. Shang, P. L. Kang, X. J. Zhang, Z. P. Liu, *Wiley Interdiscip. Rev: Comput. Mol. Sci.* **2019**, *9*, e1415.
[65] G. Kresse, J. Furthmüller, *Phys. Rev. B* **1996**, *54*, 11169.
[66] P. E. Blöchl, *Phys. Rev. B* **1994**, *50*, 17953.
[67] J. P. Perdew, K. Burke, M. Ernzerhof, *Phys. Rev. Lett.* **1996**, *77*, 3865.
[68] H. J. Monkhorst, J. D. Pack, *Phys. Rev. B* **1976**, *13*, 5188.
[69] G. Kresse, D. Joubert, *Phys. Rev. B* **1999**, *59*, 1758.

# NANO · MICRO small

## Supporting Information

Enhanced Hydrogen Evolution Catalysis of Pentlandite due to the Increases in Coordination Number and Sulfur Vacancy during Cubic-Hexagonal Phase Transition

*Yuegao Liu, Chao Cai, Shengcai Zhu, Zhi Zheng, Guowu Li, Haiyan Chen, Chao Li, Haiyan Sun, I-Ming Chou, Yanan Yu, Shenghua Mei\* and Liping Wang\**

Supporting Information

for *Small*, DOI: **10.1002/smll.202311161**

Enhanced Hydrogen Evolution Catalysis of Pentlandite due to the Increases in Coordination Number and Sulfur Vacancy during Cubic-Hexagonal Phase Transition

*Yuegao Liu†, Chao Cai‡, Shengcai Zhu§, Zhi Zheng¶, Guowu Li, Haiyan Chen, Chao Li, Haiyan Sun, I-Ming Chou, Yanan Yu, Shenghua Mei\*, Liping Wang\**

First author Email: liuyg@idsse.ac.cn

†‡§¶These authors contributed equally to this work.

\*Corresponding author Email: mei@idsse.ac.cn; wanglp3@sustech.edu.cn

**This file includes:**
    Supplementary Text (The features of the X-ray photoelectron spectroscopy data)
    Figures S1–S14
    Tables S1–S11

**Other Supplementary Materials for this manuscript include the following:**
    Movie S1

## Supplementary Text

**The features of the X-ray photoelectron spectroscopy data**

The X-ray photoelectron spectroscopy (XPS) data of cPn and hPn before and after the hydrogen evolution performance measurements were tested to determine the oxidation state of the samples (Table S9).

Overall, the peak positions and peak areas indicated by Fe 2p, Ni 2p, and S 2p before the HER (pre-HER) of cPn and hPn are essentially the same (Figures S12a, S13a, and S13c).

The intense peaks between 160.0 and 165.0 eV are assigned to sulfide S species. Four 2p doublets at 161.3 and 162.5 eV, 162.1 and 163.3 eV, 163.0 and 164.2 eV, and 163.6 and 164.8 eV are attributed to $S^{2-}$, $S_2^{2-}$, $S_n^{2-}$, and $S_n^0$, respectively.[1] The higher binding energy is deconvoluted into one doublet, at 168.5 and 169.7 eV, corresponding to the sulfate species.

Fe 2p XPS spectra show doublets at 706.4 and 719.5 and at 711.3 and 724.9 eV, representing metallic iron and oxidic iron species, respectively (Figure S13a, b). There are no hints of the presence of a shakeup satellite structure, characteristic of the $Fe^{2+}$ at approximately 714.7 eV.[1] This is consistent with the results of the Fe3p spectra (Figure S13e). Peaks at 52.8 and 56.5 eV correspond to the features of metallic iron and oxidic iron, respectively,[1–2] and no characteristic peak of $Fe^{2+}$ was observed at 53.7 eV (Figure S13e). The Ni 2p regions presented in Figure S13c-d show the presence of metallic nickel, which is represented by a doublet at 852.6 and 869.9 eV, and oxidic $Ni^{2+}$ species at 855.8 and 873.1 eV. Two shakeup satellites of oxidic $Ni^{2+}$ species likely in $Ni(OH)_2$ are present at 861.2 and 879.1 eV.[3]

The XPS features changed substantially after 10 h of HER. For cPn, compared with the pre-HER sample, the post-HER sample in the acidic solution only exhibits a weaker $S_n^0$ peak at 163.8 eV and stronger $SO_4^{2-}$ peak at 169.8 eV (Figure S12c), and the characteristic peaks of $S^{2-}$, $S_2^{2-}$, and $S_n^{2-}$ disappear.

For hPn, compared with the pre-HER sample, the post-HER sample in the acidic solution exhibits a weaker $S_2^{2-}$ peak at 162.3 eV, a stronger $S_n^0$ peak at 163.6 eV, and a stronger $SO_4^{2-}$ peak at 170.2 eV, and the characteristic peaks of $S^{2-}$ and $S_n^{2-}$ disappear (Figure S12b, d). In addition, a new peak for $SO_3^{2-}$ occurs at 166.1 eV.

For the Fe 3p region, after the HER, no metallic iron but only oxidic iron was observed (Figure S13f). Additionally, the binding energy was increased after the HER on the Fe 3p region (Figure S13e, f).

**Supplementary figures**

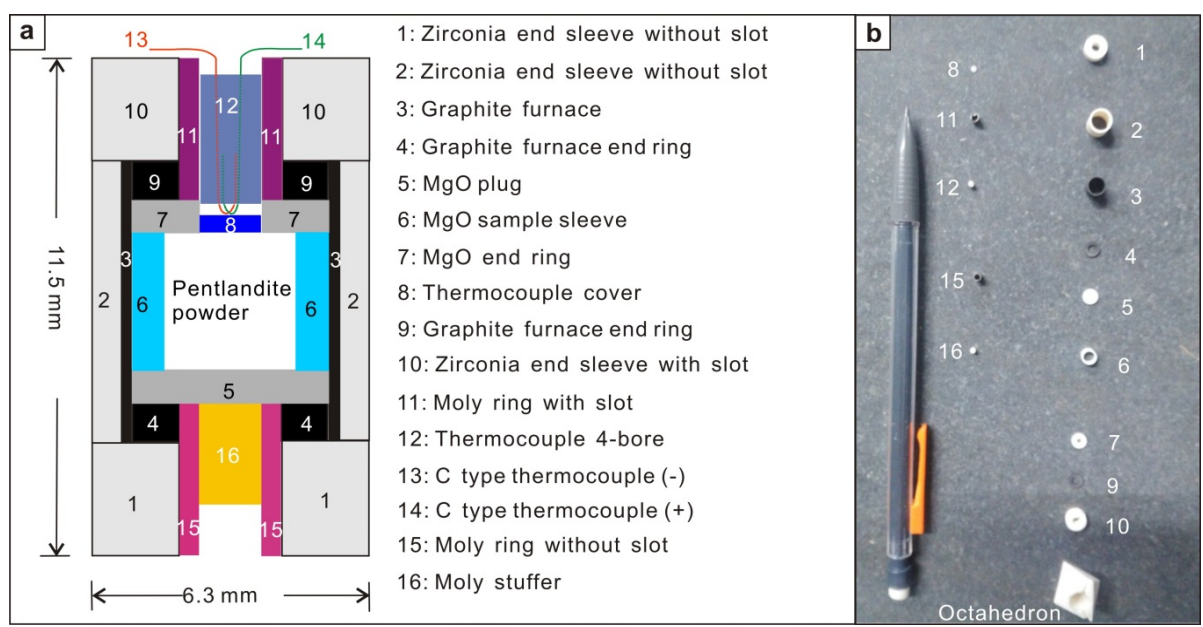

**Figure S1. a)** Scale drawings of 14/8 "G2" box-heater assembly for the production of hexagonal pentlandite. **b)** The physical map of the main components in Figure a (Ref. [4]).

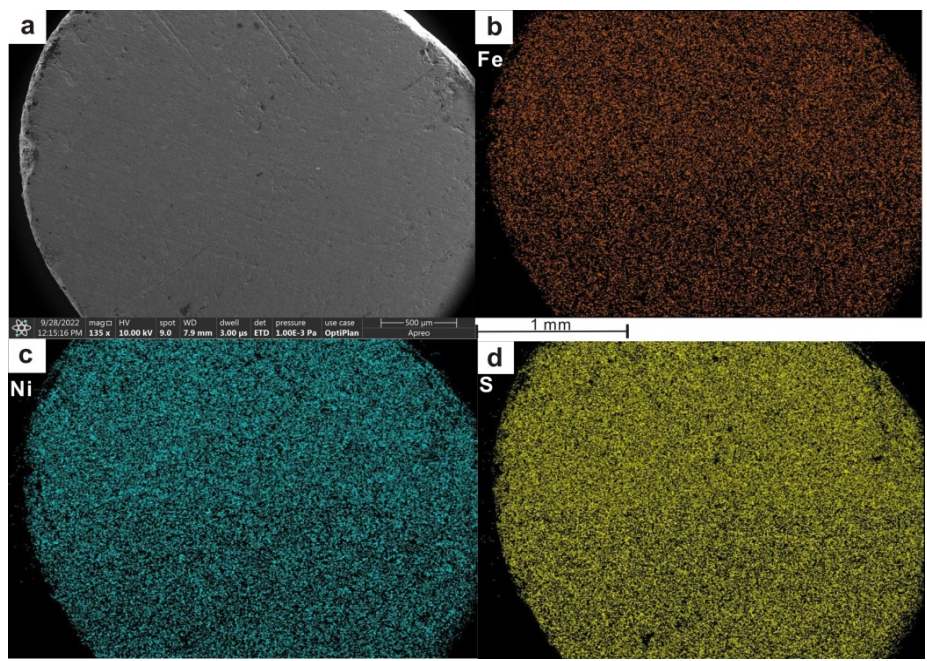

**Figure S2**. Map analyses of the synthesized hexagonal pentlandite using SEM.

    **a)** The hPn sample with a diameter of ~3 mm.

    **b)** Fe element map analyses.

    **c)** Ni element map analyses.

    **d)** S element map analyses.

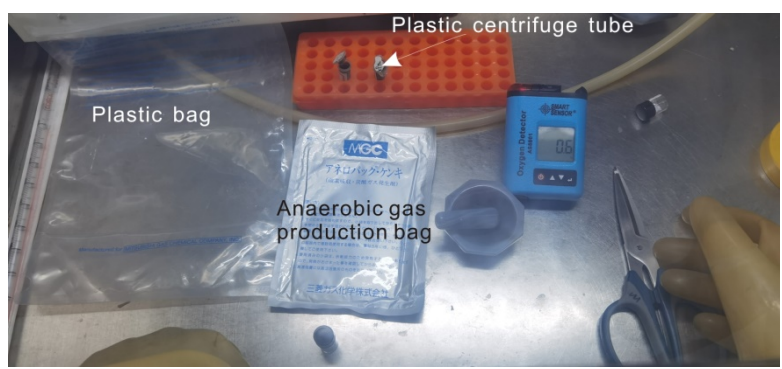

**Figure S3**. The glove box where the samples were ground and sealed with an anaerobic gas production bag placed in a plastic bag for storage.

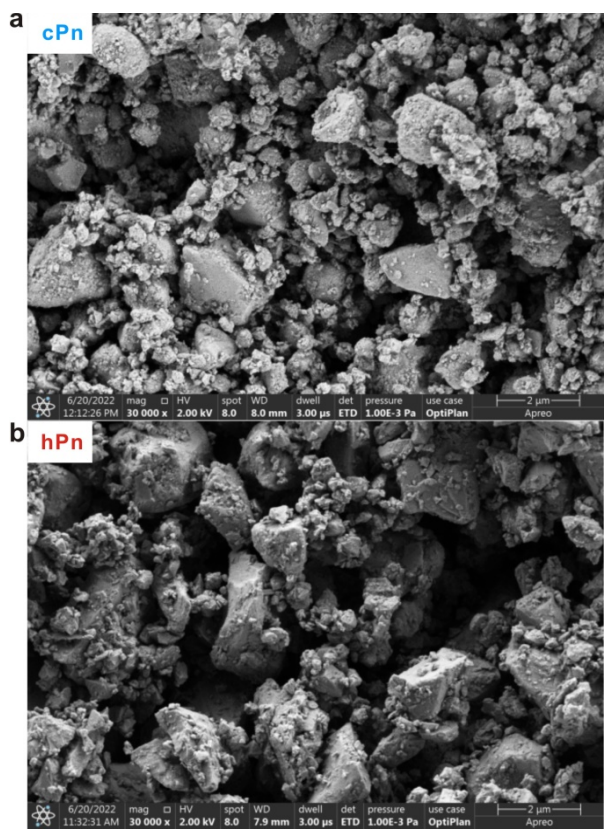

**Figure S4.** SEM image of the cubic pentlandite (Figure a) and hexagonal pentlandite (Figure b) for electrochemical testing.

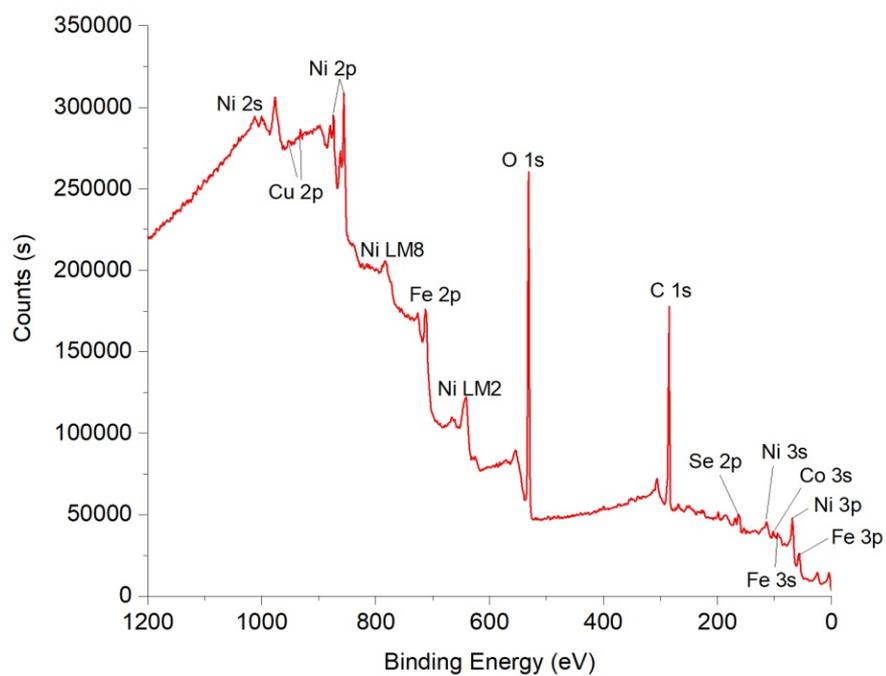

**Figure S5**. X-ray photoemission survey spectrum of hexagonal pentlandite.

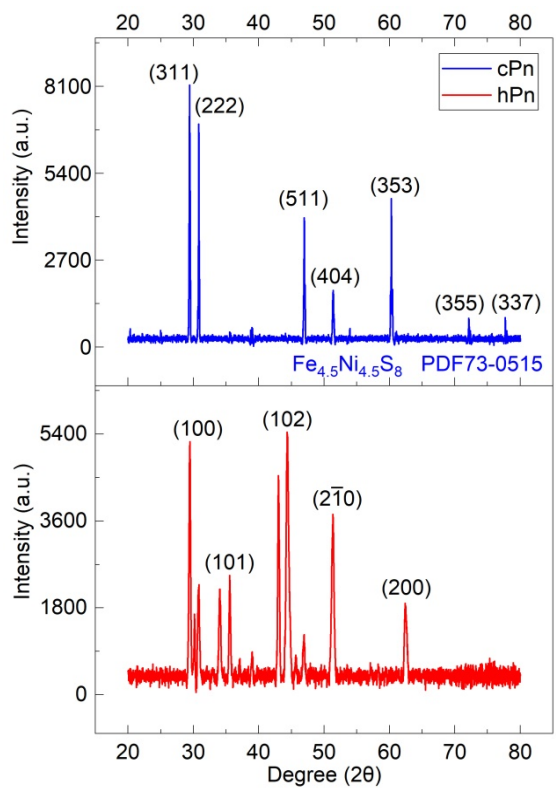

**Figure S6**. Powder XRD data of cPn and hPn.

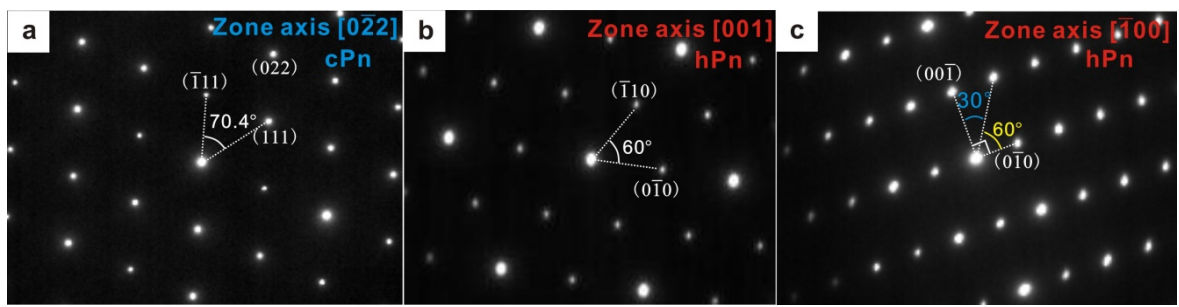

**Figure S7.** Electron diffraction patterns and crystal plane calibrations of the pentlandite.

**a)** Zone axis [0$\bar{2}$2] of the cubic pentlandite (cPn).

**b)** Zone axis [001] of the hexagonal pentlandite (hPn).

**c)** Zone axis [$\bar{1}$00] of the hexagonal pentlandite (hPn).

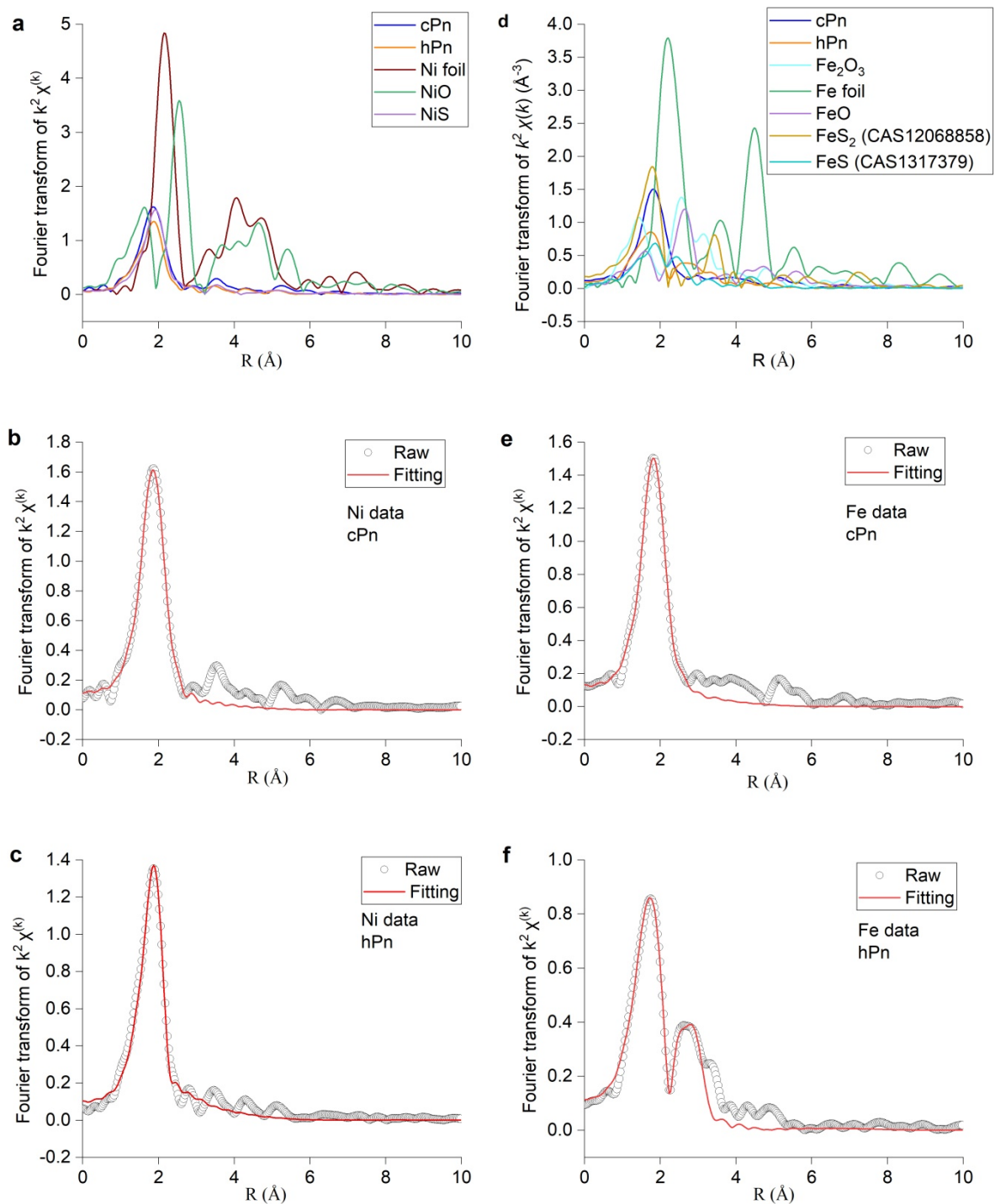

**Figure S8**. Experimental and fitting curves in real space of the Fe K-edge and Ni K-edge data. **a)** Experimental Ni K-edge data in real space. **b)** Ni K-edge fitting curves of cPn. **c)** Ni K-edge fitting curves of hPn. **d)** Experimental Fe K-edge data in real space. **e)** Fe K-edge fitting curves of cPn. **f)** Fe K-edge fitting curves of hPn.

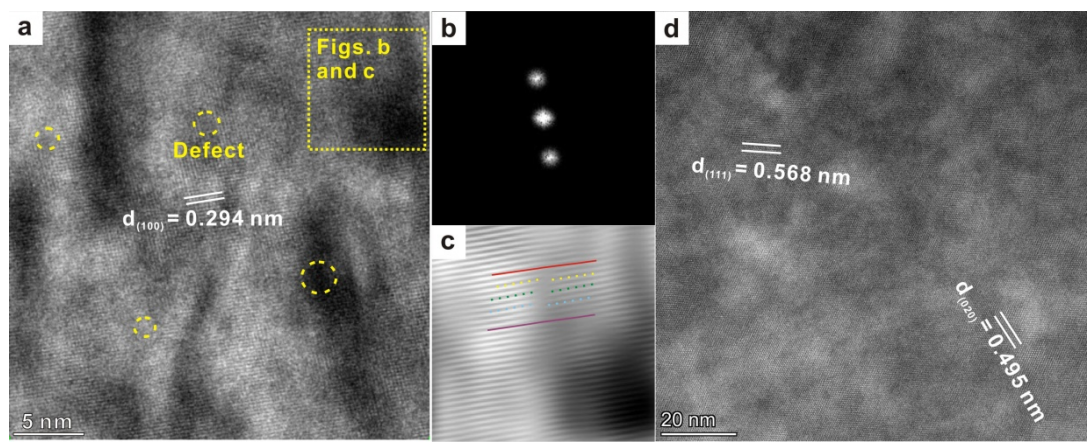

**Figure S9.** The transmission electron microscopy evidence for the enrichment of sulfur-vacancies in hexagonal pentlandite (hPn).

**a)** The high-resolution transmission electron microscopy (HRTEM) image of hPn.

**b)** Fourier transform pattern after mask.

**c)** Photo after inverse Fourier transform of Figure b.

**d)** TEM of cubic pentlandite (cPn).

.

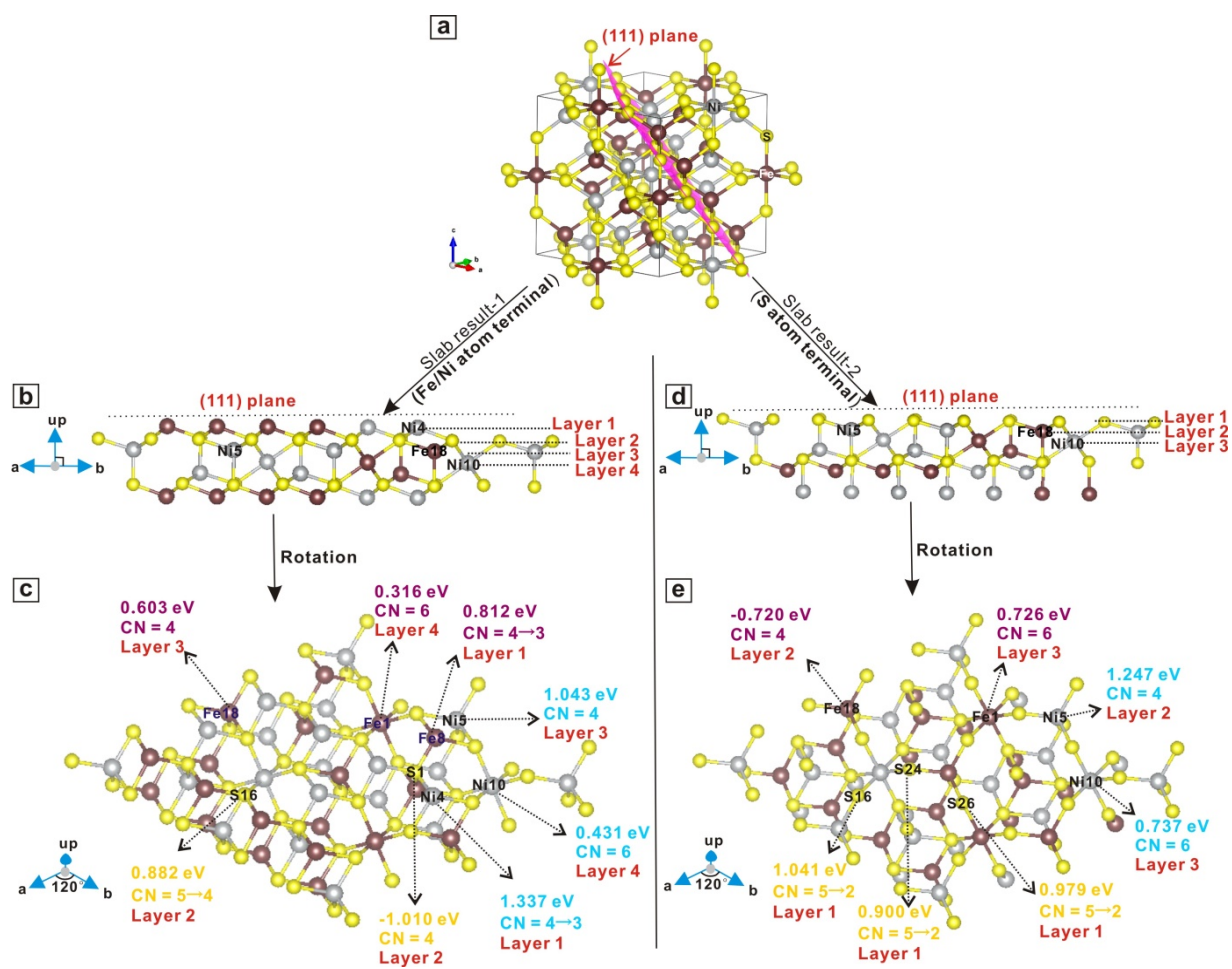

**Figure S10**. The $\Delta G_{H*}$ value of different atoms in the (111) plane of cPn.

**a)** The unit cell of cPn and the location of the Fe and Ni atoms used in this calculation.

**b)** Side view and atomic layer division of slab result-1 (Fe/Ni atom terminal).

**c)** Vertical view of slab result-1 and the $\Delta G_{H*}$ values of different atoms.

**d)** Side view and atomic layer division of slab result-2 (S atom terminal).

**e)** Vertical view of slab result-2 and the $\Delta G_{H*}$ values of different atoms. CN = coordination number; CN = X→Y means the original coordination number of the atom in the unit cell before the slab is X, but the coordination number is changed to Y after the slab.

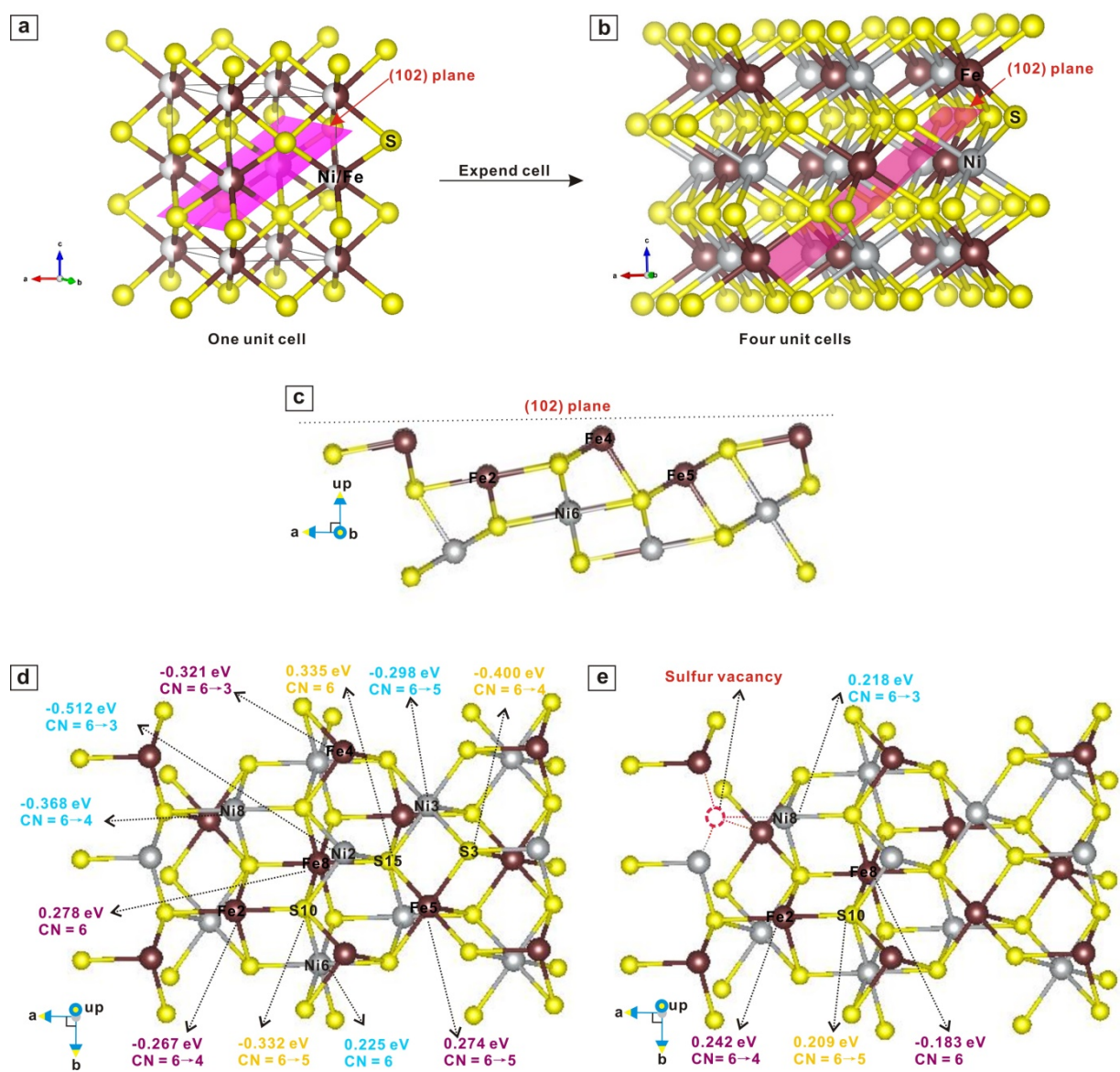

**Figure S11.** The $\Delta G_{H*}$ values of different atoms in the (102) plane of hPn.

**a)** The unit cell of hPn.

**b)** Four unit cells of hPn and the location of Fe and Ni atoms used in this calculation;

**c)** Side view of the slab result of the (102) plane.

**d)** Vertical view of the slab result of the (102) plane and the $\Delta G_{H*}$ values of different atoms.

**e)** Vertical view of the slab result of the (102) plane with one sulfur vacancy and the $\Delta G_{H*}$ values of different atoms.

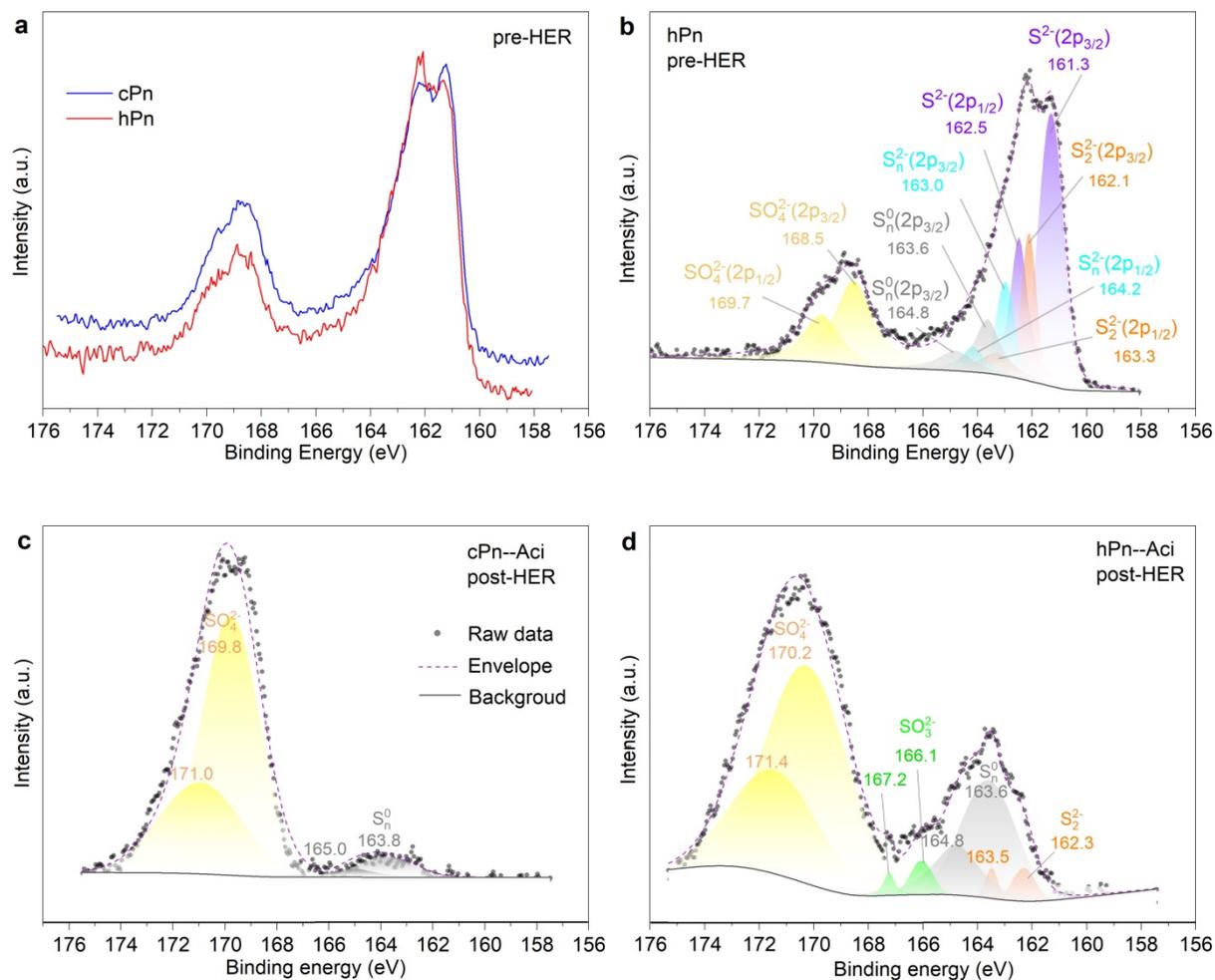

**Figure S12**. XPS Spectra of S 2p.

**a)** The S 2p XPS spectra of cPn and hPn pre-HER.

**b)** The spectrum deconvolution fits for the S 2p XPS spectra of hPn pre-HER.

**c)** The spectrum deconvolution fits for the S 2p XPS spectra of cPn post-HER in the acidic solution.

**d)** The spectrum deconvolution fits for the S 2p XPS spectra of hPn post-HER in the acidic solution.

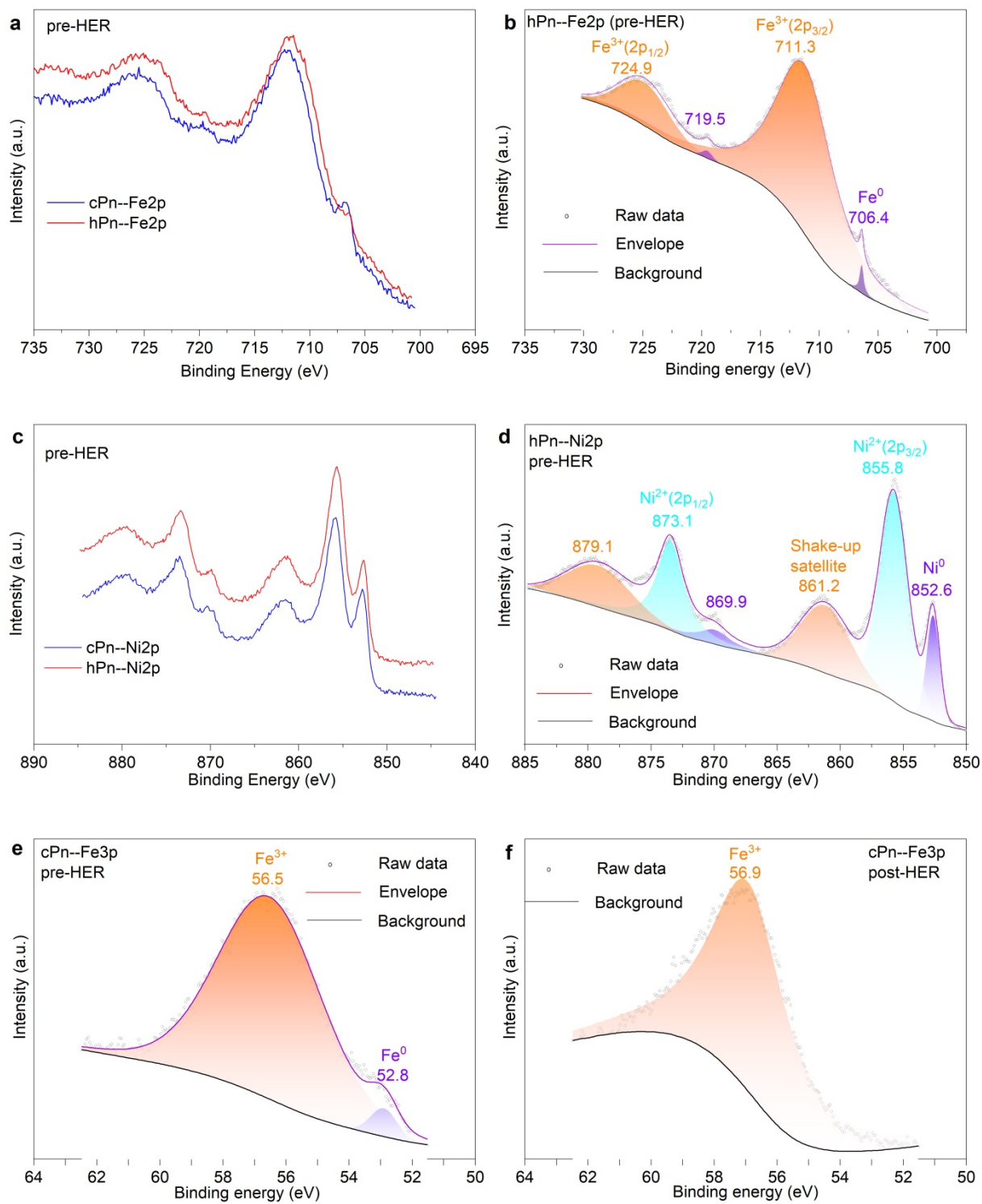

**Figure S13**. XPS Spectra of Fe 2p, Ni 2p, and Fe 3p.

**a)** XPS spectra of Fe 2p of cPn and hPn pre-HER.

**b)** The spectrum deconvolution fits for Fe 2p of hPn pre-HER.

**c)** XPS spectra of Ni 2p of cPn and hPn pre-HER.

**d)** The spectrum deconvolution fits for Ni 2p of hPn pre-HER.

**e)** The spectrum deconvolution fits for Fe 3p of cPn pre-HER.

**f)** The spectrum deconvolution fits for Fe 3p of cPn post-HER in the acidic (Aci) solution.

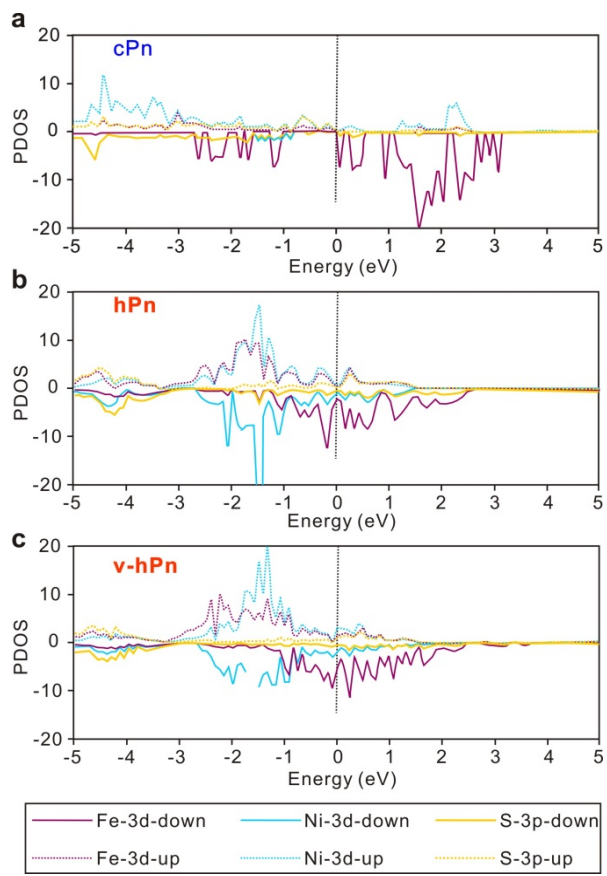

**Figure S14**. The partial density of states of cPn, hPn, and v-hPn (hPn with one sulfur vacancy).

# The title list of supplementary tables

Table S1. The composition data of cubic pentlandite (cPn) and hexagonal pentlandite (hPn) determined by EMPA (wt.%).

Table S2. Synchrotron X-ray data at different pressures and temperatures.

Table S3. X-ray photoemission survey spectra of the initial hexagonal pentlandite.

Table S4. Powder XRD data of cPn and hPn.

Table S5. EXAFS fitting parameters at the Fe and Ni K-edge for various samples.

Table S6. The Gibbs free energy of hydrogen adsorption for the typical atoms of crystal face (111) in cPn.

Table S7. The Gibbs free energy of hydrogen adsorption for the typical atoms for crystal face (102) in hPn without sulfur vacancy.

Table S8. Gibbs free energy of hydrogen adsorption for the typical atoms for crystal face (102) in hPn with sulfur vacancy.

Table S9. XPS Spectra of S 2p, Fe 2p, Ni 2p, and Fe 3p.

Table S10. EPR spectra of cPn and hPn pre-hydrogen evolution reaction (pre-HER) and post-hydrogen evolution reaction (post-HER) in the alkaline (Alk) and acidic (Aci) environments.

Table S11. The raw data for electrochemical hydrogen evolution performance.

# Supplementary movie

Movie S1. Dynamic process of phase transition from cPn to hPn

References in the supplementary

[1]     T. Yamashita, P. Hayes, *Appl. Surf. Sci.* **2008**, *254*, 2441.

[2]     A. Kochur, T. Ivanova, A. V. Shchukarev, R. Linko, A. Sidorov, M. Kiskin, V. Novotortsev, I. Eremenko, *J. Electron. Spectrosc. Relat. Phenom.* **2010**, *180*, 21.

[3]     S. Piontek, C. Andronescu, A. Zaichenko, B. Konkena, J. P. Kai, B. Marler, H. Antoni,

I. Sinev, M. Muhler, D. Mollenhauer, B. R. Cuenya, W. Schuhmann, U. P. Apfel, *ACS Catal.* **2018**, *8*, 987.

[4]    Y. Liu, I. M. Chou, J. Chen, N. Wu, W. Li, L. Bagas, M. Ren, Z. Liu, S. Mei, L. Wang, *Natl. Sci. Rev.* **2023**, 10, nwad159.